\documentclass[final,nofootinbib,aps,pre,twocolumn,showkeys,superscriptaddress,preprintnumbers,floatfix]{revtex4-1}

\usepackage{tabularx}
\usepackage{dynlearn}
\usepackage{setspace}

\colorlet{CMR_color}{black}
\colorlet{cg_color} {blue}
\colorlet{cl_color} {red}
\colorlet{ee_color} {green!66!black}

\newcommand{\naturals}{{\mathbb{N}}}

\newcommand{\semiprocess}{{state-conditioned process}}
\newcommand{\forsemiprocess}{{forward state-conditioned process}}
\newcommand{\revsemiprocess}{{reverse state-conditioned process}}
\newcommand{\physrevname}{{physical reverse representation}}
\newcommand{\formrevname}{{formal reverse representation}}

\newcommand{\classsurpname}{{class reverse surprisal}}
\newcommand{\classirrname}{{class irreversibility}}

\newcommand{\nomeftfull}{{nominal ensemble fluctuation theorem}}

\newcommand{\expeftfull}{{exponential ensemble fluctuation theorem}}

\newcommand{\Nomtcftfull}{{Nominal Class Fluctuation Theorem}}
\newcommand{\nomtcftabbr}{{NCFT}}
\newcommand{\exptcftfull}{{exponential class fluctuation theorem}}
\newcommand{\Exptcftfull}{{Exponential Class Fluctuation Theorem}}
\newcommand{\exptcftabbr}{{ECFT}}
\newcommand{\trajclassseclawfull}{{trajectory class second law}}
\newcommand{\trajclassseclawabbr}{{TCSL}}
\newcommand{\trajpartseclawfull}{{trajectory partition second law}}
\newcommand{\trajpartseclawabbr}{{TPSL}}

\newcommand{\Mst}{Z}
\newcommand{\mst}{{z}}

\newcommand{\msts}{{\mathcal Z}}

\newcommand{\Traj}{\overrightarrow \Mst}
\newcommand{\trajs}{{\overrightarrow {\mathcal{Z}}}}
\newcommand{\traj}{{\overrightarrow z}}

\newcommand{\trajat}[1]{{\overrightarrow \mst(#1)}}
\newcommand{\revtraj}{{{\overrightarrow z}^\dagger}}
\newcommand{\revtrajat}[1]{{({\overrightarrow z}^\dagger)(#1)}}

\newcommand{\dist}{\kappa}
\newcommand{\distone}{\rho}
\newcommand{\disttwo}{\sigma}

\newcommand{\distinit}{{\rho}}
\newcommand{\distinittwo}{{\sigma}}
\newcommand{\distfinal}{{\rho_\tau}}
\newcommand{\eqat}[1]{{\pi_{#1}}}
\newcommand{\eqdist}{{\pi}}

\newcommand{\Senv}{{S_\text{env}}}
\newcommand{\ssys}{{s_\text{sys}}}
\newcommand{\Entlike}{{\Sigma}}

\newcommand{\E}{E}
\newcommand{\Eat}[1]{{E_{#1}}}
\newcommand{\prot}{\overrightarrow \lambda}

\newcommand{\Feq}{F^\text{eq}}

\newcommand{\f}{f}

\newcommand{\metregion}{m}

\newcommand{\forP}{{P}}
\newcommand{\forp}{{p}}

\newcommand{\physrevp}{{r'}}
\newcommand{\physrevP}{{R'}}
\newcommand{\rev}[1]{{#1^\dagger}}

\newcommand{\revP}{{R}}
\newcommand{\revp}{{r}}

\newcommand{\classes}{\mathcal{C}}
\newcommand{\class}{C}
\newcommand{\classtwo}{C'}

\newcommand{\partition}{{Q}}

\newcommand{\avgat}[2]{{\left\langle{#1}\right\rangle_{#2}}}
\newcommand{\avgatens}[1]{{\avgat{#1}{\trajs}}}
\newcommand{\avgatclass}[1]{{\avgat{#1}{\class}}}

\newcommand{\classsurp}{{\Theta}}
\newcommand{\classsurpat}[1]{{\classsurp_{#1}}}
\newcommand{\classsurpatclass}{{\classsurpat{\class}}}
\newcommand{\classirr}{{\psi}}
\newcommand{\classirrat}[1]{{\classirr_{#1}}}
\newcommand{\classirratclass}{{\classirrat{\class}}}

\newcommand{\avgirr}{{\Psi}}
\newcommand{\avgirrat}[1]{{\avgirr_{#1}}}
\newcommand{\avgirratclass}{{\avgirrat{\class}}}
\newcommand{\avgavgirratpart}{{\mathbf{\Psi}}_\partition}

\begin{document}

\def\ourTitle{Trajectory Class Fluctuation Theorem
}

\def\ourAbstract{The Trajectory Class Fluctuation Theorem (TCFT) substantially strengthens the
Second Law of Thermodynamics---that, in point of fact, can be a rather weak
bound on resource fluxes. Practically, it improves empirical estimates of free
energies, a task known to be statistically challenging, and has connected
the microscopic dynamics with the mesoscopic information processing
in experimentally-implemented Josephson-junction information engines. The
development here justifies that empirical analysis, explicating its
mathematical foundations.\\
The TCFT reveals the thermodynamics induced by macroscopic system
transformations for each measurable subset of system trajectories. In this, it
directly combats the statistical challenge of extremely rare events that
dominate thermodynamic calculations. And, it reveals new forms of free
energy---forms that can be solved analytically and practically estimated.
For engineered systems, it provides a toolkit for diagnosing the
thermodynamics responsible for system functionality.
Conceptually, the TCFT unifies a host of previously-established fluctuation
theorems, interpolating from Crooks' Detailed Fluctuation Theorem (single
trajectories) to Jarzynski's Equality (full trajectory ensembles).
}

\def\ourKeywords{integral fluctuation theorem, detailed fluctuation theorem, free energy
}

\hypersetup{
  pdfauthor={Gregory Wimsatt},
  pdftitle={\ourTitle},
  pdfsubject={\ourAbstract},
  pdfkeywords={\ourKeywords},
  pdfproducer={},
  pdfcreator={}
}

\author{Gregory Wimsatt}
\email{gwwimsatt@ucdavis.edu}
\affiliation{Complexity Sciences Center and Physics and Astronomy Department, University of California at Davis, One Shields Avenue, Davis, CA 95616}

\author{Alexander B. Boyd}
\email{abboyd@ucdavis.edu}
\affiliation{Complexity Sciences Center and Physics Department, University of California at Davis, One Shields Avenue, Davis, CA 95616}

\author{James P. Crutchfield}
\email{chaos@ucdavis.edu; Corresponding author}
\affiliation{Complexity Sciences Center and Physics Department, University of California at Davis, One Shields Avenue, Davis, CA 95616}

\date{\today}
\bibliographystyle{unsrt}

\title{\ourTitle}

\begin{abstract}
	\ourAbstract
\end{abstract}



\preprint{\arxiv{2207.03612}}

\title{\ourTitle}
\date{\today}
\maketitle

\setstretch{1.1}


\section{Introduction}

The century-old study of thermodynamic fluctuations was rejuvenated with the
discovery of fluctuation theorems in the very late twentieth century
\cite{Evan93a, Gall95a, Jarz97a, Croo99a, Jarz11a}. Jarzynski's Equality
\cite{Jarz97a} and Crooks' Detailed Fluctuation Theorem \cite{Croo98a}, in
particular, have been used to infer and relate thermodynamic properties of
small systems driven by transformations very far from equilibrium. Fluctuation
theorems revealed that stochastic deviations from equilibrium in small-scale
systems obey specific functional forms. That is, fluctuations are lawful.

In fact, the equilibrium Second Law and its nonequilibrium generalization can
be derived from the stronger equalities provided by fluctuation theorems
\cite{Jarz11a}. In addition, equilibrium and nonequilibrium free energy changes
are readily obtained from those same stronger equalities \cite{Jarz97a,
Jarz04a}. And, this allows estimating free energies given sufficient sampling
of a thermodynamic process. This has been carried out successfully for RNA and
DNA configurational free energies \cite{Liph02a, Mara08a, Juni09a} and quantum
harmonic oscillators \cite{Deff08a}. However, obtaining free energies is
generally quite challenging statistically due to the existence of rare events
that dominate the exponential average work \cite{Jarz06a, Asba17a}.

Many of these results rely on averaging thermodynamic quantities over
trajectory ensembles. We recently introduced the \emph{trajectory class
fluctuation theorem} (TCFT) that focuses instead on subsets of
trajectories---trajectory \emph{classes} \cite{Wims19a}. While a restricted
form of the TCFT had been noted previously \cite{Asba17a}, we present a theorem
that applies to arbitrary measurable subsets of trajectories. And, we derive a
suite of results that lift prior limitations. Namely, by considering
information about one or more trajectory classes, we markedly strengthen
statements of the Second Law \cite{Wims19a}. Practically, too, by using
trajectory classes with high probability, we overcome limitations in estimating
free-energy differences due to finite sampling.

The TCFT introduces a new level of flexibility central to extracting free
energy differences in a wide variety of empirical settings. The \emph{detailed
fluctuation theorem} (DFT) \cite{Croo98a}, the basis from which these results
are derived, relies on comparing state trajectory probabilities evaluated from
a \emph{forward experiment} and and a \emph{reverse experiment} to evaluate the
entropy production in the forward experiment. However, the DFT's predictive
capacity is severely hampered by the fact that the state trajectories are
typically so numerous that their individual probabilities are extremely small,
if not zero. It is then virtually impossible to sample sufficient data to
reliably estimate those probabilities. Moreover, it is rare in an experiment to
have complete information about a system trajectory.

To meet these challenges, Ref. \cite{Wims19a}'s TCFT provides a practical
computational advantage by estimating entropy from a much smaller space of
trajectory classes---classes that can be tailored to specific experimental data
and constraints. The following further expands on the TCFT's experimental relevance by generalizing to the case in which the reverse experiment does not
necessarily start in a special distribution---i.e., the distribution conjugate
to the ending distribution of the forward experiment. This generalization
requires introducing a new thermodynamic quantity known as the \emph{entropy difference}, which can be interpreted as entropy production in special cases,
but has important thermodynamic consequences regardless. To illustrate, we
apply the TCFT to metastable processes---processes where the system begins and
ends in metastable distributions---and show how to derive metastable free
energies with an appropriately initialized set of experiments \cite{Parr15a}.

Theoretically, the TCFT unites a wide variety of prior fluctuation theorems.
These include theorems that range systematically from Crooks' DFT
\cite{Croo99a} to Jarzynski's \emph{integral fluctuation theorem} (IFT)
\cite{Jarz97a}. That noted, the TCFT itself can be derived from broader
theorems still \cite{Croo99b, Seif12a}; see App. \ref{app:AltTCFTDerivations}.
The TCFT's strength then is in its balance of specificity and generality.

Developing the TCFT proceeds as follows. Section \ref{sec:background} presents
fluctuation theorem building blocks, culminating in Crooks' (DFT) and the basic
IFTs. Section \ref{sec:tcft} uses the DFT to introduce and prove the TCFT.
Section \ref{sec:tsl} shows how it strengthens the Second Law in light of
process data. However, the Second Law and its strengthened forms are much more
useful when the free-energy difference is known so that bounds on work can be
established. And so, Sec. \ref{sec:const_f} shows how to use the TCFT to solve
for free-energy differences beyond just the equilibrium free-energy difference.
Section \ref{sec:tyranny} demonstrates how the TCFT overcomes the tyranny of
rare events when estimating free energies from data. Section
\ref{sec:related_results} highlights related results and surveys how the TCFT
encapsulates them. Finally, we briefly discuss several subtle aspects of
applying the TCFT in Sec. \ref{sec:discussion}, considering the application to
a recent nanoscale flux qubit experiment \cite{Wims19a}. Section
\ref{sec:conclusion} concludes.

\section{Background}
\label{sec:background}

\subsection{Model, Probability Densities, and Time Reversal}
\label{sec:model}

Consider a system interacting with both a control device and a thermal
environment over a time interval $[0,\tau]$ from time $0$ to time $\tau$. The
device enacts a control protocol $\prot$ over the time interval to influence
the system. Specifically, at any time $t$, $\prot(t)$ specifies the function
from system state to system energy, called the \emph{energy landscape}, at time
$t$. The protocol therefore both affects how the system evolves and requires
energy, which we call \emph{work}, to be exchanged between the system and the
control device. The thermal environment has inverse temperature $\beta$ so that
energy, denoted \emph{heat}, flows between the system and environment. No other
interactions exist. Altogether, this induces a stochastic dynamic in the system
as it evolves from time $0$ to time $\tau$.

We model time as either continuous or discrete. In general, the resultant set
$T$ of times is some subset of the interval $[0,\tau]$ that includes $0$ and
$\tau$.

We model the system states as either its microstates or as some coarse-graining
of its microstates. Denote a particular system \emph{state} as $\mst$ and a
particular system state \emph{trajectory} as $\traj$, with $\trajat{t}$ the
realized system state at time $t\in T$. We let $\msts$ denote the set of all
possible states and $\trajs$ the set of all possible trajectories.

The energy of the system in state $\mst$ at time $t$ is denoted
$\Eat{t}(\mst)$. The energy change of the system over an entire trajectory
$\traj$ is then:
\begin{align*}
  \Delta\E(\traj)=\Eat\tau(\trajat{\tau})-\Eat{0}(\trajat{0})
  ~.
\end{align*}
We designate positive work or heat to mean that energy flowed
into the system from the control device or thermal environment, respectively.

We require that the net work $W(\traj)$ is a function of trajectory. This can
be achieved if $T$ and $\msts$ are sufficiently refined. (For example,
$T=[0,\tau]$ and $\msts$ is the system's set of microstates.) Then, by
conservation of energy, heat $Q(\traj)$ must also be a function of trajectory
and we have the First Law for each trajectory:
\begin{align*}
  \Delta\E(\traj)=W(\traj)+Q(\traj)
  ~.
\end{align*}

We describe probabilities of states with state probability densities, which we
refer to as \emph{distributions}. These are functions of $\msts$ that when
integrated over a region of state space give the corresponding probability of
occupying that region. As an important example, the system's equilibrium
distribution $\eqat{t}$ for energy landscape $\Eat{t}$ is the corresponding
Boltzmann distribution:
\begin{align*}
\eqat{t}(\mst) = e^{-\beta(\Eat{t}(\mst)-\Feq_t)}
  ~,
\end{align*}
where $\Feq_t$ is the system's equilibrium free energy at time $t$ and
$\beta$ is the inverse temperature as denoted in statistical mechanics. Note
that if $\msts$ is discrete, then a state probability density evaluated at a
state is in fact the corresponding probability of that state. That is,
integration over discrete spaces can simply be taken to be summation.

Similarly, we describe probabilities of trajectories with trajectory
probability densities. These densities are functions of $\trajs$ that, when
integrated over a region of trajectory space, give the corresponding
probability of a trajectory occupying that region.

If the state space $\msts$ is a Euclidean space of dimension $d$, as is typical
for spaces of microstates, integration over $\msts$ can of course be done with
a $d$-dimensional Riemannian integral. However, the trajectory space $\trajs$
is much too large to be Euclidean when the set of times $T$ is continuous.
This then requires a more powerful notion of integration.

A solution can be found via measure theory but we treat the subject only
briefly here. (A sequel provides the details \cite{Wims24a}.) A measure is a function on the regions of a space that returns an
amount of some ``quantity'', such as probability, in a region. We choose a
particular measure on trajectory space and refer to it as a \emph{base measure}.
A probability density is then a function that, when Lebesgue-integrated
via the base measure over a region of space, yields the corresponding
probability of that region.
Defining appropriate base measures on a continuous-time trajectory space is
rather technical, though, and so we leave the discussion to the sequel.

We specify the dynamic induced by a protocol $\prot$ via a set of trajectory
probability densities, one for each possible initial state $\mst$. This
gives the probability density of evolving the system trajectory $\traj$
conditioned on starting in a state $\mst$:
\begin{align*}
  \Pr_{\prot}(\Traj=\traj|\Mst_0=\mst)
  ~,
\end{align*}
where $\Traj$ and $\Mst_0$ are random variables for the trajectory and the
initial state, respectively. Call such a set of trajectory
probability densities a \emph{\semiprocess}.

For system state $\mst$, the \emph{time-reverse state} $\rev \mst$,
or simply \emph{reverse state}, is the same state with all components
odd under time reversal flipped in sign. (Recall momentum or spin.) For
state trajectory $\traj$, $\revtraj$ is the \emph{reverse state
trajectory}: $\revtrajat{t} = \rev {(\trajat{\tau-t})}$ for
$0 \leq t \leq \tau$. If $\dist$ is a distribution over system states,
then $\rev \dist$ is the reverse distribution, defined by
$(\rev \dist)(\mst) = \dist(\rev \mst)$. Note that time reversal of a
state, trajectory, or distribution is an involution,
meaning that time reversal acted twice on any such object returns the original
object.

For a given protocol $\prot$, consider the corresponding \emph{time-reverse
protocol} $\rev\prot$. $\prot$ dictates a set of forces and fields that are
applied to the system as a function of time. Enacting $\rev\prot$ then requires
applying these same influences but in the reverse order as well as flipping the
sign of time-odd influences, such as magnetic fields. Time reversing is
therefore also involutional on protocols.

For simplicity when working with time-reversal, we require:
\begin{itemize}
  \item $\tau-t$ to be in $T$ for each $t$ in $T$ and
  \item $\rev\mst$ to be in $\msts$ for each $\mst$ in $\msts$.
\end{itemize}
These basic symmetry requirements are satisfied in typical models in
statistical mechanics and nonequilibrium thermodynamics.

\subsection{Forward and Reverse: Experiments and Processes}

The main objects of study are a system's \emph{forward} and \emph{reverse
processes}, which result from forward and reverse experiments. The
\emph{forward experiment} consists of an initial distribution $\distinit$ and
the forward control protocol $\prot$, which evolves the distribution over the
time interval $(0,\tau)$. Similarly, for some state distribution $\disttwo$,
the \emph{reverse experiment} applies the reverse control protocol
$\rev{\prot}$ to the initial distribution $\disttwo^\dagger$. We refer to
$\distinit$ and $\disttwo$ as \emph{privileged distributions}, emphasizing that
different choices for these distributions result in different predictions given
by the TCFTs and other fluctuation theorems.

Through the control protocol $\prot$, the forward experiment produces the
\emph{\forsemiprocess}\, $\forp$, where we denote the probability density of
a trajectory $\traj$ conditioned on the initial state $\mst$ as:
\begin{align*}
  \forp(\traj | \mst)
  \equiv \Pr_{\prot} (\Traj = \traj | \Mst_0 = \mst)
  ~,
\end{align*}
Similarly, the \emph{\revsemiprocess}\, $\revp'$ is obtained under the
reverse protocol $\rev{\prot}$:
\begin{align*}
  \physrevp(\traj | \mst)
  \equiv \Pr_{\rev \prot} (\Traj = \traj | \Mst_0 = \mst)
  ~.
\end{align*}

We suppose throughout that microscopic reversibility holds for the system.
This means that when the system evolves along trajectory $\traj$, the
environment's net entropy change is:
\begin{align}
\Delta \Senv(\traj) & = -\beta Q(\traj) \nonumber \\
  & = \ln \frac{\forp(\traj | \trajat{0})}
  {\physrevp(\revtraj | \rev{(\trajat{\tau})})}
 ~.
\label{eq:MR}
\end{align}
Recall that microscopic
reversibility can be derived, for example, under Markov \cite{Croo98a},
Hamiltonian \cite{Jarz00a, Jarz06a}, or Langevin \cite{Seif05} assumptions.

The \emph{forward process} is a trajectory probability density $\forP$
specified by the initial distribution $\distinit$ and the \forsemiprocess.
Under $\forP$, the probability density of a trajectory $\traj$ is:
\begin{align*}
\forP(\traj) &\equiv \distinit(\trajat{0}) \forp(\traj | \trajat{0}) ~.
\end{align*}

For any time $t$, we marginalize $\forP$ to find the evolved state distribution
$\distone_t$. For each region $A$ in system state space, let $\forP_t(A)$ be
the probability of the system's state being in $A$ at time $t$, and let
$\class$ be the set of trajectories that occupy region $A$ at time $t$. Then
$\distone_t$ is the state probability density that satisfies:
\begin{align*}
  \forP_t(A)
  &=\int_A d\mst\,\distone_t(\mst) \\
  &=\int_\class d\traj \forP(\traj)
  ~.
\end{align*}

Analogously, the \emph{reverse process} is a trajectory probability density
$\physrevP$ determined by the initial distribution $\rev\disttwo$ and the
\revsemiprocess.  Under $\physrevP$, the probability density of a trajectory
$\traj$ is:
\begin{align}
\physrevP(\traj) &\equiv (\rev\disttwo) (\trajat{0})
  \physrevp(\traj | \trajat{0}) ~.
 \label{eq:physrevP}
\end{align}

To simplify the following, we use an alternate representation for the reverse
process---the \emph{\formrevname} $\revP$---another trajectory probability
density.  For each trajectory $\traj$, we define:
\begin{align}
  \revP(\traj) \equiv \physrevP(\revtraj)
  ~.
  \label{eq:revP}
\end{align}
To keep the two representations distinct, the original representation
$\physrevP$ of the reverse process is called the \emph{\physrevname}.

\subsection{Detailed Fluctuation Theorem}

Applying the principle of microscopic reversibility to forward and reverse
processes leads directly to a \emph{detailed fluctuation theorem} (DFT). First,
define the \emph{system state entropy}, or \emph{system state surprisal}, of a
given state $\mst$ for a given distribution $\dist$ \cite{Seif05} as:
\begin{align*}
\ssys(\mst; \dist) = -\ln\dist(\mst)
 ~.
\end{align*}
Second, define the \emph{system entropy difference} for a trajectory $\traj$
in terms of the two privileged distributions $\rho$ and $\sigma$:
\begin{align}
\Delta \ssys(\traj) \nonumber
  &\equiv \ssys(\trajat{\tau}; \disttwo) - \ssys(\trajat{0}; \distone) \\
  &= \ln\frac{\distone(\trajat{0})}{\disttwo(\trajat{\tau})}
 ~.
\label{eq:dssys} 
\end{align}
This is similar, but more general than the \emph{change in system entropy}
\cite{Seif05} of the forward experiment. The latter is the difference in
surprisal of the system in the forward experiment:
\begin{align*}
\ln \frac{\distinit(\trajat{0})}{\distfinal(\trajat{\tau})}
  ~.
\end{align*}
The system entropy difference is the change in system entropy if the
reverse experiment is initialized in the time reversal of the final
distribution of the forward experiment, meaning $\disttwo=\rho_\tau$.

We designate the \emph{entropy difference} as the difference in system
state entropy and the change in environmental entropy:
\begin{align}
\Entlike(\traj) &= \Delta \ssys(\traj) + \Delta \Senv(\traj)
 ~.
\label{eq:entlike_def}
\end{align}
Again, this is similar, but more general than another familiar quantity.
When we choose $\disttwo=\rho_\tau$, then $\Entlike(\traj)$ gives
the entropy change of the system plus that of the environment for $\traj$,
which is known as the \emph{entropy production} of the forward experiment.

Together in Eq. (\ref{eq:entlike_def}), Eqs. (\ref{eq:MR}) and (\ref{eq:dssys})
yield an expression for the entropy difference:
\begin{align}
\Entlike(\traj)
& = \ln \frac{\distinit(\trajat{0}) \forp(\traj | \trajat{0})}
  {\distinittwo(\trajat{\tau})\physrevp(\revtraj | \rev{(\trajat{\tau})})}
  ~.
\label{eq:EntDiffSub}
\end{align}
The numerator is $\forP(\traj)$.  And, by Eqs. (\ref{eq:physrevP}) and
(\ref{eq:revP}):
\begin{align*}
  \revP(\traj) &= (\rev \disttwo) ((\rev{\traj})(0))
  \physrevp(\rev\traj | (\rev\traj)(0) ) \\
  & = (\rev \disttwo) (\rev{(\trajat{\tau})})
  \physrevp(\rev\traj | \rev{(\trajat{\tau})}) \\
  &= \disttwo(\trajat{\tau}) \physrevp(\rev\traj | \rev{(\trajat{\tau})})
  ~.
\end{align*}
These two observations translate Eq. (\ref{eq:EntDiffSub}) into a DFT:
\begin{align}
  \Entlike(\traj)
  &= \ln \frac{\forP(\traj)}{\revP(\traj)}
 ~.
\label{eq:gdft1}
\end{align}
This is a fluctuation theorem obtained previously in a Langevin setting
\cite{Seif05} and a generalization of the Crooks fluctuation theorem
\cite{Croo98a} to the case of arbitrary privileged distributions.

Continuing in this way, we introduce a constraint on the forward and reverse
processes:
\begin{align}
  \label{eq:support_condition}
  \revP(\traj)=0 ~\text{when}~ \forP(\traj)=0
  ~.
\end{align}
The motivation is that the forward process needs to ``cover'' all the
trajectories that are significant to the reverse process. The failure of
Condition (\ref{eq:support_condition}) generally introduces subprobabilistic
measures and densities for the reverse process that complicate the development
and, in any case,  may not be experimentally accessible. When considering the
TCFT applied to a trajectory class $\class$, described shortly in Sec.
\ref{sec:tcft}, such complications are avoided so long as Condition
(\ref{eq:support_condition}) holds for all $\traj\in\class$. For simplicity of
discussion, we assume it holds for all $\traj\in\trajs$.

Assuming that the heat $Q(\traj)$ is finite for all $\traj$, then microscopic
reversibility Eq. (\ref{eq:MR}) guarantees
$\revp(\rev\traj|\rev{(\trajat\tau)})=0$ wherever $\forp(\traj|\trajat0)=0$.
Then Condition (\ref{eq:support_condition}) is met so long as
$\distinittwo(\trajat\tau)=0$ for any $\traj$ where $\distinit(\trajat0)=0$.
The most straightforward way to ensure this is to let $\distinit$ have at
least a small amount of probability density on all system states.
Note that if $\Eat0$ is everywhere finite, then $\eqat0$ has full support.

\subsection{Work and Free Energy}
\label{sec:ent_diff_decomps}

The following shows that, for any given trajectory, the entropy difference
decomposes into the requisite work in the forward experiment minus a
\emph{difference in nonequilibrium free energy}. Note that the latter is more
general than the change in nonequilibrium free energy of the forward
experiment. This realization yields important versions of the fluctuation
theorems. In particular, it allows extracting the change of free energy---an
important privileged-distribution-dependent but protocol-independent
quantity---from the work.

To see this, first define the \emph{state free energy} for a distribution
$\dist$ and system energy function $\E$:
\begin{align*}
\f(\mst; \dist, \E) = \E(\mst) + \beta^{-1} \ln \dist(\mst)
 ~.
\end{align*}
An important example occurs when $\dist$ is the equilibrium distribution for
$\E$.  In that case, the state free energy is constant over all $\mst$
and is the equilibrium free energy.

Second, define the \emph{trajectory free-energy difference} for the forward and
reverse processes as:
\begin{align*}
  \Delta \f(\traj)
  &\equiv \f(\trajat{\tau}; \distinittwo, \Eat{\tau})
    - \f(\trajat{0}; \distinit, \Eat{0}) \\
  &= \Delta \E(\traj)
  + \beta^{-1} \ln \frac{\distinittwo(\trajat{\tau})}{\distinit({\trajat{0}})}
   ~.
\end{align*}
Again, the latter echoes a familiar thermodynamic quantity: the \emph{change in
nonequilibrium free energy} $\Delta F^\text{neq}$ for the forward experiment
\cite{Parr15a}. The free-energy difference reduces to the nonequilibrium free
energy change when $\disttwo=\rho_\tau$.

Using the first law---$\Delta E(\traj)
= W(\traj) + Q(\traj)$---rewrite the entropy difference in terms of the work
and free-energy difference:
\begin{align}
  \label{eq:entlike_decomp}
  \Entlike(\traj)
  &= -\ln \frac{\distinittwo(\trajat{\tau})}{\distinit(\trajat{0})}
    - \beta Q(\traj) \nonumber \\
  &= \beta \big(\Delta \E(\traj) - \Delta \f(\traj) - Q(\traj)\big)
  \nonumber \\
&= \beta \big( W(\traj) - \Delta \f(\traj) \big)
   ~,
\end{align}
And so, the entropy difference is the work in the forward experiment minus the
difference in nonequilibrium free energy.

\subsection{Ensemble Fluctuation Theorems and the Second Law}

From Eq. (\ref{eq:gdft1})'s DFT, it is easy to derive two general fluctuation
theorems. First, there is the \nomeftfull:
\begin{align}
  \avgatens{\Entlike}
  &= -\int d\traj \forP(\traj) \ln \frac{\revP(\traj)}{\forP(\traj)}
  \nonumber \\
  &= \DKL{\forP}{\revP}_\trajs
  ~.
\label{eq:nomeft}
\end{align}
Here, $\avgatens{\cdot}$ denotes an ensemble average over all trajectories
$\trajs$. And, $\DKL{\forP}{\revP}_\trajs$ is the Kullback-Leibler divergence
between the forward and reverse process taking all trajectories $\trajs$ as
argument.

The divergence tracks the mismatch between the distributions. Generally, it is
nonnegative and vanishes only when the distributions are equal over all events.
In the present case, the ensemble average entropy difference is zero only when
$\forP(\traj) = \revP(\traj)$ for all $\traj \in \trajs$.

The divergence's nonnegativity is tantamount to a generalized Second
Law of thermodynamics---one that bounds average entropy differences:
\begin{align}
  \avgatens{\Entlike} \geq 0
  ~.
  \label{eq:esl}
\end{align}
This includes the familiar bound on entropy production, but different
choices of the privileged distributions lead to new bounds on thermodynamic
quantities.

Applying Eq. (\ref{eq:entlike_decomp}), the average free energy change bounds the work done in the forward experiment:
\begin{align}
\avgatens{W} &\geq \avgatens{\Delta\f} \nonumber \\
  &= \int d\traj \forP(\traj) \Delta\f(\traj) \nonumber \\
  &= \int d\traj \forP(\traj) \f(\trajat{\tau};\disttwo,E_\tau) \nonumber \\
  & \qquad
   - \int d\traj \forP(\traj) \f(\trajat{0};\distinit,E_0) \nonumber \\
  &= \int \!\!  d\mst \distfinal(\mst) \f(\mst;\disttwo,E_\tau)
  \! - \!\!  \int\!\!   d\mst \distinit(\mst) \f(\mst;\distinit,E_0)
  .
\label{eq:wf_bound}
\end{align}
The average work is determined by the forward process exclusively and
therefore must not have any actual dependence on the second privileged
distribution $\disttwo$, despite the latter's appearance in the first term on
the RHS. And yet, the bound must hold for whichever distribution is chosen for
$\disttwo$.

This begs the question, for which $\disttwo$ is Eq. (\ref{eq:wf_bound})
tightest? The answer is the final-time distribution $\distfinal$:
\begin{align*}
  \int d\mst \distfinal(\mst) \f(\mst; \disttwo, E_\tau)
  &= \int d\mst \distfinal(\mst) [E_\tau(\mst) + \beta^{-1} \ln \disttwo(\mst)] \\
  &= \int d\mst \distfinal(\mst) [E_\tau(\mst) + \beta^{-1} \ln \distfinal(\mst) \\
   &\quad - \beta^{-1} \ln \distfinal(\mst) + \beta^{-1} \ln \disttwo(\mst)] \\
  &= \int d\mst \distfinal(\mst) \f(\mst; \distfinal, E_\tau) \\
   &\quad - \beta^{-1} \DKL{\distfinal}{\disttwo} \\
  &\leq \int d\mst \distfinal(\mst) \f(\mst; \distfinal, E_\tau)
  ~,
\end{align*}
where:
\begin{align*}
  \DKL{\distfinal}{\disttwo}
  &= \int d\mst \distfinal(\mst) \ln \frac{\distfinal(\mst)}{\disttwo(\mst)}
\end{align*}
is nonnegative. Therefore, the free-energy difference is generally less than
the change in nonequilibrium free energy, which gives the strongest bound on
work production:
\begin{align}
  \label{eq:wf_strictest_bound}
  \avgatens{W} & \geq \avgatens{\Delta F^\text{neq}}
  \geq \avgatens{\Delta\f}
  ~.
\end{align}

Even though the nonequilibrium free-energy change for the forward process
provides the tightest bound on work when it is used in Eq. (\ref{eq:wf_bound}),
there are other useful alternatives. This flexibility is helpful as it may be
difficult to determine the precise final-time distribution $\distfinal$. Or, we
may be more interested in the system after it relaxes to its equilibrium state
$\eqat{\tau}$ determined by the final-time energy function $E_\tau$:
\begin{align*}
\int d\mst \distfinal(\mst) \f(\mst; \eqat{\tau}, E_\tau)
  &= \int d\mst \distfinal(\mst) (E_\tau(\mst) \\
  &\quad + \beta^{-1}\ln[e^{\beta(E_\tau(\mst)-\Feq_\tau)}]) \\
  &= \Feq_\tau
  ~.
\end{align*}
If we start the system in equilibrium $\distinit = \eqat{0}$, a similar
calculation shows that, from Eq. (\ref{eq:wf_bound}):
\begin{align*}
\avgatens{W} \geq \Delta \Feq
  ~,
\end{align*}
with:
\begin{align*}
  \Delta \Feq = \Feq_\tau - \Feq_0
  ~.
\end{align*}
Thus, this gives the equilibrium Second Law which applies to systems that start
in equilibrium, but end in arbitrary distributions close to or far from
equilibrium. And so, it also applies without using any information about
the final distribution of the forward experiment $\distfinal$.

This all said, the assumption of equilibrium is rather restrictive. It is often
possible to set more informed bounds on work invested by using incomplete
information about the initial distribution $\rho_0$ and final distribution
$\rho_\tau$. Such distributions are metastable, but provide a convenient method
of improving work production estimates. Section
\ref{sec:metastable_process_wb} discusses this shortly.

Finally, from the DFT we can obtain the \expeftfull:
\begin{align}
  \avgatens{e^{-\Entlike}}
  &=\int d\traj\forP(\traj)\frac{\revP(\traj)}{\forP(\traj)} \nonumber \\
  & =\int d\traj\revP(\traj) \nonumber \\
  &= 1
  ~.
\label{eq:expeft}
\end{align}

Again assuming equilibrium privileged distributions, the equilibrium
free-energy difference can be extracted from the average:
\begin{align*}
  \avgatens{e^{-\Entlike}}
  &= \avgatens{e^{-\beta(W - \Delta\Feq)}} \\
  & = \avgatens{e^{-\beta W}} e^{\beta\Delta\Feq}
  ~,
\end{align*}
giving Jarzynski's Equality:
\begin{align}
  \avgatens{e^{-\beta W}} = e^{-\beta \Delta \Feq}
  ~.
\label{eq:jarzynski}
\end{align}
Remarkably, this allows extracting equilibrium free-energy differences from
work statistics of highly nonequilibrium processes. However, the exponential
free-energy difference $\Delta\f$ cannot generally be extracted from
$\avgatens{e^{-\beta(W-\Delta\f)}}$ for any choice of $\distinit$ besides
$\eqat{0}$. Additionally, estimating even the free-energy difference from
experiment using Eq. (\ref{eq:jarzynski}) can lead to sampling issues due to
rare but resource-dominant events. We use the TCFT in Secs.
\ref{sec:const_f} and \ref{sec:tyranny} to confront these two problems.

\section{Trajectory Class Fluctuation Theorem}
\label{sec:tcft}

The preceding results on nonequilibrium thermodynamic processes are statements
concerning either individual trajectories or ensemble averages---that is,
averages over all trajectories. As we will see, though, a markedly broader
picture emerges when considering averages that lie between. Specifically, the
following treats arbitrary subsets of trajectories, called \emph{trajectory
classes}, as the main players in analyzing fluctuations. The resulting
\emph{trajectory class fluctuation theorem} (TCFT) reveals relationships
involving probabilities of trajectory classes, averages conditioned on
trajectory classes, and thermodynamic quantities of interest. And, the results
include strengthened versions of the Second Law, solving for general free
energies from works, and statistically efficient methods for finding those free
energies from data. Additionally, the TCFT provides a general form that
subsumes many fluctuation theorems.

The section begins by introducing trajectory classes and relevant relationships
and probabilities involving them. Then it establishes the TCFT from these
building blocks. It ends with a discussion of the TCFT's scope, suggesting a
way to treat zero-probability classes and providing an example use.

\subsection{Trajectory Classes}

Every trajectory class is a subset of $\trajs$ for which the forward and
reverse processes assign probability. Formally, the set of all trajectory
classes $\classes$ must constitute a $\sigma$-algebra over the trajectories
$\trajs$. However, to stay with the physics, the following does not focus on
measure-theoretic details. (The sequel focuses on the latter.) Instead, we
first discuss several example classes and thereafter interpret any subset of
$\trajs$ of practical interest to be part of the assumed $\sigma$-algebra and,
therefore, to be a trajectory class.

Clearly, the nature of $\msts$ and $T$, such as whether these sets are
discrete or continuous, must determine the form of the trajectories and
therefore determine the form of the trajectory classes.
However, many intuitive types of trajectory classes exist quite broadly, as illustrated by the following list of Examples:
\begin{enumerate}
      \setlength{\topsep}{-1pt}
      \setlength{\itemsep}{-1pt}
      \setlength{\parsep}{-1pt}
\item All trajectories that at a given time $t\in T$ are in a
	specified finite volume of state space,
\item All trajectories that at a given time $t\in T$ are in a
	particular state,
\item All trajectories that have an entropy difference in a specified finite
	range of values,
\item All trajectories that have a particular value of the
	entropy difference,
\item All trajectories: $\trajs$, and
\item The singleton $\{\traj\}$ for any trajectory $\traj\in\trajs$.
\end{enumerate}
The singleton trajectory classes of Example (6) provides one instance where the
model of the system determines whether a trajectory class exists. For a finite
number of times $T$, we can assume the singleton trajectory classes exist.
However, for technical reasons, such trajectory classes often fail to exist for
continuous-time processes.
See Sec. \ref{sec:related_results} for more examples as they apply to known
results.

\subsection{Trajectory Class Quantities}

For each trajectory class $\class$, we denote the forward and reverse process
probabilities as $\forP(\class)$ and $\revP(\class)$, respectively.
They are given by:
\begin{align*}
  \forP(\class) = \int_\class d\traj \forP(\traj)
  \quad \text{and} \quad
  \revP(\class) = \int_\class d\traj \revP(\traj)
  ~.
\end{align*}

To derive the TCFT for a trajectory class $\class$, we require that
$\forP(\class)$ be nonzero. However, classes like Examples (2), (4), and (6)
above will often have zero probability. Section \ref{sec:zero_probs} discusses
the use of the TCFT in such cases. Until then, we will assume that
$\forP(\class) \neq 0$.

The forward and reverse class-conditioned trajectory probability
densities are, for $\traj \in \class$:
\begin{align*}
  \forP(\traj | \class) \equiv \frac{\forP(\traj)}{\forP(\class)}
  \quad \text{and} \quad
  \revP(\traj | \class) \equiv \frac{\revP(\traj)}{\revP(\class)}
  ~,
\end{align*}
respectively. The class-conditioned densities vanish
for $\traj \notin \class$. When $\revP(\class)=0$, we allow
$\revP(\traj | \class)$ to be any probability.

We also make frequent use of class-conditioned expectation values:
\begin{align*}
  \langle f \rangle_\class
  \equiv \int d\traj \forP(\traj | \class) f(\traj)
  ~,
\end{align*}
for arbitrary functions $f$ of $\trajs$.

The reverse of a trajectory class $\class$ is defined as:
\begin{align*}
  \rev\class
  &= \{\rev\traj | \traj\in\class\}
  ~.
\end{align*}
Then the physical reverse probability $\physrevP(\rev\class)$ of obtaining
trajectory class $\rev\class$ during the reverse experiment is equal to the
formal reverse probability $\revP(\class)$:
\begin{align*}
  \revP(\class)=\physrevP(\rev\class)
  ~.
\end{align*}
With the above equality, we can then obtain an empirical estimate of
$\revP(\class)$ from reverse experiment data.

We then define two important quantities for any class. The
\emph{\classsurpname} measures how much more surprising an occurrence of class
$\class$ is in the reverse process than in the forward process:
\begin{align*}
  \classsurpatclass
  \equiv \ln\frac{\forP(\class)}{\revP(\class)}
  ~.
\end{align*}
While $\classsurpatclass$ does not have explicit dependence on any specific
trajectory $\traj$, the \emph{\classirrname} $\classirratclass$ does:
\begin{align*}
  \classirratclass(\traj)
  \equiv \ln\frac{\forP(\traj | \class)}{\revP(\traj|\class)}
  ~.
\end{align*}

\subsection{Fluctuation Theorem}

We now introduce two fluctuation theorems that arise from the preceding
setup.  Given their close relation, together they constitute the TCFT.

The \classsurpname\, and \classirrname\, form a key decomposition of the
entropy difference for $\traj\in\class$:
\begin{align}
  \Entlike(\traj) &= \ln\frac{\forP(\traj)}{\revP(\traj)} \nonumber \\
&= \ln\frac{\forP(\class)\forP(\traj|\class)}
    {\revP(\class)\revP(\traj|\class)} \nonumber \\
&= \classsurpatclass + \classirratclass(\traj)
  \label{eq:entlike_classdecomp}
  ~.
\end{align}
Equation (\ref{eq:entlike_classdecomp}) can fail in some cases. For, example if
$R(\class)=0$ then it fails when the trajectory is outside of the class, which allows nonzero trajectory probability $\revP(\traj)\neq0$.
But a trajectory $\traj\in\class$ for which Eq.
(\ref{eq:entlike_classdecomp}) fails must occur with zero probability in the
forward process. So, any instance of $\Entlike$ can be substituted with
$\classsurpatclass+\classirratclass$ in any class-conditioned average.
To derive the TCFT, we will only use $\Entlike$ in class-conditioned averages
and so we will treat Eq. (\ref{eq:entlike_classdecomp}) as valid in all cases.

The \classirrname\, $\classirratclass$ takes two important forms. The first
when averaged directly; the second when averaging its exponential. When class
averaging directly, we obtain:
\begin{align}
\avgirratclass & \equiv \avgatclass{\classirratclass} \nonumber \\
  & = \DKL{\forP}{\revP}_\class
  ~,
\label{eq:classirravg}
\end{align}
where:
\begin{align}
\DKL{\forP}{\revP}_\class
  &\equiv \int d\traj \forP(\traj|\class)
  \ln\frac{\forP(\traj|\class)}{\revP(\traj|\class)}
  ~,
\end{align}
is the class-conditioned divergence between $\forP$ and $\revP$. It is a
nonnegative quantity, being a Kullback-Leibler divergence, that measures how
closely the reverse process emulates the forward process when conditioned on
the class $\class$.

Directly class averaging the entropy difference of Eq.
(\ref{eq:entlike_classdecomp}) gives the following.

\begin{The}
  \Nomtcftfull\, (\nomtcftabbr):
  For any trajectory class $\class$ where $\forP(\class)\neq0$:
\begin{align}
  \label{eq:nomtcft}
  \avgatclass{\Entlike} &= \classsurpatclass + \avgirratclass \\
&= \ln\frac{\forP(\class)}{\revP(\class)}
+ \DKL{\forP}{\revP}_\class
  \nonumber
  ~.
\end{align}
\end{The}

When Sec. \ref{sec:tsl} considers refinements of the Second Law, this equality
proves its worth in describing the average entropy difference while
conveniently isolating the precise trajectory information into the nonnegative
\classirrname.

Turning now to exponential class averages, we have a fruitful identity:
\begin{align*}
  \avgatclass{e^{-\classirratclass}}
  &= \int d\traj \forP(\traj|\class)
  \frac{\revP(\traj | \class)}{\forP(\traj | \class)} \\
  &= \int d\traj \revP(\traj | \class) \\
  &= 1
  ~.
\end{align*}
And, Eq. (\ref{eq:entlike_classdecomp}) gives:
\begin{align*}
  \avgatclass{e^{-\Entlike}}
  &= \avgatclass{e^{-\classsurpatclass-\classirratclass}} \\
  & = e^{-\classsurpatclass}\avgatclass{e^{-\classirratclass}}
  ~.
\end{align*}
Combining these yields the following.

\begin{The}
  \Exptcftfull\, (\exptcftabbr):
\begin{align}
  \label{eq:exptcft}
  \langle e^{-\Entlike} \rangle_{\class} &= e^{-\classsurpatclass} \nonumber \\
  & = \frac{\revP(\class)}{\forP(\class)}
   ~.
  \end{align}
\label{thm:ECFT}
\end{The}

The equality's significance lies in relating the entropy difference to the
rather simple \classsurpname\, without any possibly-detailed specification of
trajectory probabilities beyond the class probabilities. We use it, shortly, to
develop straightforward equalities about entropy difference, work, free energy
changes, and forward and reverse class probabilities.

\subsection{Scope: Nonzero Probability Classes}
\label{sec:nonzero_probs}

The TCFT spans a large collection of fluctuation theorems.
By choosing $\class$
to be all possible trajectories $\trajs$, we obtain the ensemble fluctuation
theorems. That is:
\begin{align*}
\classsurpat{\trajs} &= \ln\frac{\forP(\trajs)}{\revP(\trajs)} \\
   & = 0,
\end{align*}
and:
\begin{align*}
\avgirrat{\trajs}
&= \DKL{\forP}{\revP}_{\trajs}
  ~.
\end{align*}
So that:
\begin{align*}
\avgat{\Entlike}{\trajs} & = \DKL{\forP}{\revP}_{\trajs} \text{~and} \\
  \avgat{e^{-\Entlike}}{\trajs}
  & = e^{-0} \\
  & = 1
~.
\end{align*}
These recover the ensemble fluctuation theorems of Eqs. (\ref{eq:nomeft}) and
(\ref{eq:expeft}).

Choosing any proper subset $\class$ of $\trajs$ identifies a new set of
FTs---the TCFT applied to the more refined class $\class$. We will consider
several types of classes in Secs. \ref{sec:const_f} and \ref{sec:tyranny}. Also
see Sec. \ref{sec:related_results} for examples from the literature.

So consider the opposite extreme, letting $\class=\{\traj\}$ consist of a
single trajectory $\traj\in\trajs$. We require $\forP(\{\traj\})>0$ in order
to apply the TCFT, but note that $\forP(\{\traj\})$, the probability of
obtaining the particular trajectory $\traj$ in the forward process, is
typically zero for continuous-state or continuous-time processes.
Keep in mind that $\forP(\{\traj\})$ is distinct from the
probability density $\forP(\traj)$.
Then:
\begin{align*}
\classsurpat{\{\traj\}} &= \ln\frac{\forP(\{\traj\})}{\revP(\{\traj\})}
\end{align*}
and:
\begin{align*}
  \avgirrat{\{\traj\}}
&= \DKL{\forP}{\revP}_{\{\traj\}} \\
  &= 0 ~.
\end{align*}
And, so:
\begin{align*}
  \avgat{\Entlike}{\{\traj\}}
  &= \ln\frac{\forP(\{\traj\})}{\revP(\{\traj\})}
  ~.
\end{align*}
Integrating $\forP(\traj)$ over $\{\traj\}$ yields $\forP(\{\traj\})$,
so $\forP(\{\traj\})=\forP(\traj)d\traj$.
Similarly, $\revP(\{\traj\})=\revP(\traj)d\traj$, meaning:
\begin{align*}
  \avgat{\Entlike}{\{\traj\}}
  &= \ln\frac{\forP(\traj)}{\revP(\traj)}
  ~.
\end{align*}
And:
\begin{align*}
  \avgat{\Entlike}{\{\traj\}}
  &=\int_{\{\traj\}}d\traj'\forP(\traj'|\{\traj\})\Entlike(\traj') \\
  &=\Entlike(\traj)
  ~.
\end{align*}
From the above equalities, we recover the DFT as expressed in Eq. (\ref{eq:gdft1}).

\subsection{Scope: Zero Probability Classes}
\label{sec:zero_probs}

For a sufficiently large trajectory space $\trajs$, such as with continuous-time
processes, the vast majority of singleton classes $\{\traj\}$ must have
zero probability.  This is because $\trajs$ is then uncountable and only
countably many disjoint events can have nonzero probability.  In general, many
classes of interest, like those whose trajectories with a specific work
value, will have zero probability and so will not be directly subject to the
TCFT.

Fortunately, we can still apply the TCFT less directly to a class
$\class$ such that $\forP(\class)=0$.
One method has practical appeal.
Consider a second trajectory class $\class ' \supset \class$ that is nearly
identical to $\class$
except for extensions in some dimensions of trajectory space such that
$\forP(\class ') > 0$. For example, if $\class$ is all trajectories that
pass through a specific state $\mst$ at a particular time, one might let
$\class'$ be all trajectories that pass through a small but
nonzero-probability neighborhood of states surrounding $\mst$ at that time. Then, one
considers the TCFT applied to the broadened class $\class'$ in place of the
original class $\class$. Since it is necessary that
experimentally-sampled classes have nonzero probability in any case, this
approach is attractive. The remaining art is to choose and use an appropriate
alternative class $\class'$ for the class of interest $\class$.

Carrying this further, a second possibility suggests itself. One that is
more satisfying theoretically and yields results involving probability
densities. Consider a limiting procedure that applies the TCFT to smaller and
smaller classes containing a class of interest. To give one such scheme,
consider a trajectory class $\class$ and a sequence of classes
$\class_1, \class_2, \ldots$ such that:
\begin{itemize}
      \setlength{\topsep}{-1pt}
      \setlength{\itemsep}{-1pt}
      \setlength{\parsep}{-1pt}
  \item $ \class_1 \supseteq \class_2 \supseteq \ldots $ ,
  \item $ \bigcap_{n\in\naturals} \class_n = \class $ , and
  \item $\forP(\class_n) > 0$ for all $n$
  ~.
\end{itemize}
Then consider Eqs. (\ref{eq:nomtcft}) and (\ref{eq:exptcft}) for each
$\class_n$ and limiting behavior as $n \rightarrow \infty$ to evaluate entropy differences for classes with zero probability.

As an example, consider the class $\class$ of trajectories with work value
$\widetilde W$ generated by a process:
\begin{align*}
  \class = \{\traj | W(\traj) = \widetilde W\}
  ~.
\end{align*}
If the work distribution's values are continuous for the process, then each
particular work value has zero probability of occurring. However, consider a
class $\classtwo$ that allows a range of works:
\begin{align*}
  \classtwo
  = \{\traj | \widetilde W - \epsilon < W(\traj) < \widetilde W + \epsilon \}
  ~,
\end{align*}
for some $\epsilon>0$.
Such a class generically has nonzero probability and can be used in place
of $\class$.

Considering Thm. \ref{thm:ECFT}'s \exptcftabbr\, with equilibrium privileged
distributions, we have:
\begin{align*}
  \avgat{e^{-\beta(W - \Delta\Feq)}}{\classtwo}
  &= \frac{\revP(\classtwo)}{\forP(\classtwo)}
  ~.
\end{align*}
As $\epsilon$ decreases, the work distribution for $\classtwo$ necessarily
narrows, so that $e^{-\beta(W - \Delta\Feq)}$ approaches $e^{-\beta(\widetilde
W - \Delta\Feq)}$. And, if the work distributions for the forward and reverse
processes are continuous functions of work value, then
$\revP(\classtwo)$ and $\forP(\classtwo)$ will eventually shrink at the same,
constant rate. In this case, $\revP(\classtwo) / \forP(\classtwo)$ converges
to a ratio of work densities at $\widetilde W$.

This procedure recovers Crooks' work fluctuation theorem \cite{Croo99a}:
\begin{align}
  \label{eq:crooks_work}
  e^{-\beta(W - \Delta\Feq)} = \frac{\revP(W)}{\forP(W)}
  ~,
\end{align}
where $\forP(W)$ and $\revP(W)$ are the probability densities of obtaining work
$W$ in the forward and reverse processes, respectively. Note that for a
trajectory $\traj$ with work $W$, the work under the reverse protocol
$\rev\prot$ and reverse trajectory $\rev\traj$ is $-W$.  So
$\revP(W)=\physrevP(-W)$.

In fact, the TCFT introduced here can be strengthened to directly address
classes of zero probability without the need for approximations or limiting
schemes.  This strengthening is done with the measure-theoretical notion of
conditional expectation.  However, an exposition requires a more thorough
treatment of measure theory and so we treat it in the sequel.

\section{Strengthening the Second Law}
\label{sec:tsl}

Having established the TCFT and outlined how it subsumes existing fluctuation
theorems, the following turns attention to bounds on the entropy difference
that are similar to but stronger than the Second Law. For these, we need
only to determine, experimentally or computationally, the probabilities of
trajectories in the forward and reverse processes. First, we find a Second Law
for individual trajectory classes. Second, this yields a fluctuation theorem
involving multiple trajectory classes that together partition all trajectories.
This fluctuation theorem then produces the
\emph{Trajectory Partition Second Law} that
sets a strictly stronger bound on the ensemble-average entropy difference than
the traditional Second Law.

\subsection{Trajectory Class Second Law}
\label{sec:tcsl}

Discarding the class-average class irreversibility $\avgirratclass$ in Eq.
(\ref{eq:nomtcft})---a nonnegative quantity---gives the
\emph{\trajclassseclawfull} (\trajclassseclawabbr):
\begin{align}
\label{eq:tcsl}
\avgatclass{\Entlike} &\geq \classsurpatclass \nonumber \\
  & = \ln \frac{\forP(\class)}{\revP(\class)}
~.
\end{align}
Thus, the \classsurpname\, $\classsurpatclass$---a quantity that only depends
on $\forP(\class)$ and $\revP(\class)$---
bounds the class averaged entropy difference
$\avgatclass{\Entlike}$.

This is similar to Eq. (\ref{eq:esl})'s Second Law that bounds the ensemble-averaged entropy difference to be nonnegative. However, Eq. (\ref{eq:tcsl}) is
more precise since it uses more information---the class probabilities in the
forward and reverse processes. Section \ref{sec:tpsl} elaborates on this
advantage over the ensemble Second Law.

Equation (\ref{eq:nomtcft}) says that the average entropy difference
$\avgatclass{\Entlike}$ is close to the \classsurpname\, $\classsurpatclass$
when the class-average \classirrname\, $\avgirratclass$ is small. Appendix
\ref{sec:app_classirr} shows that having such a small \classirrname\, is
equivalent to the class $\class$ having a narrow entropy-difference
distribution.

\subsection{Trajectory Partition Fluctuation Theorem}

Now partition all trajectories $\trajs$ into trajectory classes forming a
collection $\partition$. Then averaging Eq. (\ref{eq:nomtcft}) over
all classes in $\partition$ gives an equality obtained in Ref. \cite{Gome08a}
that is generalized to arbitrary privileged distributions:
\begin{align}
  \langle \Entlike \rangle_{\trajs}
  &= \sum_{\class \in \partition} \forP(\class)
    \langle \Entlike \rangle_{\class} \nonumber \\
  &= \sum_{\class \in \partition} \forP(\class)
    (\classsurpatclass + \avgirratclass) \nonumber \\
&= \avgat{\classsurp}{\partition}
    + \avgavgirratpart
\label{eq:gtcft_partition_avg}
  ~,
\end{align}
where the \emph{partition averaged \classsurpname} is:
\begin{align*}
  \avgat{\classsurp}{\partition}
  &\equiv \sum_{\class \in \partition} \forP(\class) \classsurpatclass \\
  &= \DKL{\forP}{\revP}_\partition
  ~.
\end{align*}
This is a Kullback-Leibler divergence over classes in the partition:
\begin{align}
  \DKL{\forP}{\revP}_\partition
  &\equiv\sum_{\class\in\partition} \forP(\class)
  \ln\frac{\forP(\class)}{\revP(\class)}
  ~.
\end{align}
And, where the \emph{partition-class averaged \classirrname} is:
\begin{align*}
  \avgavgirratpart
  &\equiv \sum_{\class \in \partition}
    \forP(\class) \avgirratclass \\
  &= \sum_{\class \in \partition} \forP(\class)
\DKL{\forP}{\revP}_\class
  ~,
\end{align*}
a weighted sum of divergences. So, the ensemble average entropy difference
decomposes into the mismatch between forward and reverse class probabilities
plus the mismatch between specific forward and reverse trajectory
probabilities in a class, averaged over all classes.

\subsection{Trajectory Partition Second Law}
\label{sec:tpsl}

Since divergences are nonnegative, Eq. (\ref{eq:gtcft_partition_avg}) leads
directly to the \emph{\trajpartseclawfull\,} (\trajpartseclawabbr) by
discarding the partition-class averaged \classirrname\, $\avgavgirratpart$:
\begin{align}
\avgat{\Entlike}{\trajs}
& \geq \avgat{\classsurp}{\partition} \nonumber \\
  & = \DKL{\forP}{\revP}_\partition
\label{eq:tpsl}
 ~,
\end{align}
where, notably, the RHS leaves out detailed trajectory information, relying
only on class probabilities.  One can use this expression to bound the entropy
difference of a system from a wide array of limited observations of the
system.  This includes coarse graining time \cite{Riec20, Wims21b} and system
state space \cite{Parr15a, Riec20, Wims21b}, as well as many other
possibilities \cite{Rold12, Mart19, Skin21}.

The information that is left out in going from Eq.
(\ref{eq:gtcft_partition_avg}) to Eq. (\ref{eq:tpsl}) is the class
irreversibility averaged over all trajectories in a class and then averaged
over all classes in the partition. So, in accordance with Sec. \ref{sec:tcsl},
if the classes are chosen to have narrow entropy-difference distributions, the
\classsurpname s will tightly bound the ensemble average entropy difference.
Specifically, Eq. (\ref{eq:tpsl}) is a tight-bound.

Contrast this with Eq. (\ref{eq:esl})'s ensemble Second Law. Since it only
states that the average entropy difference $\avgat{\Entlike}{\trajs}$ is
nonnegative, the \trajpartseclawfull\, always provides a nonnegative
improvement over the Second Law in estimating ensemble-average entropy
differences.

To emphasize, Eq. (\ref{eq:tpsl})'s \trajpartseclawabbr\, can be made
arbitrarily tight by considering finer and finer partitions $\partition$ whose
classes have narrower and narrower entropy-difference distributions,
independent of the process and how poorly Eq. (\ref{eq:esl}) bounds the average
entropy difference.

That said, partitioning trajectories into finer classes complicates solving for
and relating class probabilities. Ideally, there is middle ground with a
relatively simplified partition composed of classes that carry sufficient
information about the trajectory probabilities to tightly-bound the average
entropy difference. Reference \cite{Wims19a} provides a compelling example of
the experimental usefulness of this result, as it uses naturally defined
classes to obtain strong work estimates for trajectories in experimentally
implemented bit erasure in a flux qubit.

\section{Free Energies via Constant-Difference Classes}
\label{sec:const_f}

Paralleling the bounds on work production derived from the Second Law, Eq.
(\ref{eq:entlike_decomp}) converts the bounds of Sec. \ref{sec:tsl} into
statements involving works and trajectory free-energy differences. Thus,
determining a bound on the average work required over an arbitrary trajectory
class requires obtaining the trajectory free-energy difference. The following
shows how to find these free-energy differences given access to the work
statistics for the forward experiment and trajectory class statistics for both
the forward and reverse experiment.  Our only requirement is that the
trajectories in the class all have the same free-energy difference. The
resulting trajectory class free energy differences provide average work bounds
for a wide variety of processes. We then consider an important type of
nonequilibrium process---a metastable process---for an example application.

\subsection{Constant Free-Energy Differences}
\label{sec:const_f_theory}

For any pair of forward and reverse \semiprocess es defined by a forward
protocol $\prot$, a choice of equilibrium privileged distributions
$\distinit=\eqat0$ and $\distinittwo=\eqat\tau$ ensures a constant free-energy
difference $\Delta\f(\traj)$ over all trajectories $\traj$: the equilibrium
free energy change $\Delta\Feq$.
For nonequilibrium privileged distributions, $\distinit\neq\eqat0$ or
$\distinittwo\neq\eqat\tau$, the free-energy difference
varies over trajectories.
And, this precludes extraction of the free-energy difference from the
exponential average entropy difference as was done to obtain
Eq. (\ref{eq:jarzynski}) for $\Delta\f=\Delta\Feq$.
By focusing on trajectory classes of constant free-energy difference, though,
we can actually extract these free-energy differences from class averages.

Suppose every trajectory $\traj$ in class $\class$ has the same free energy
difference $\Delta \f_\class=\Delta\f(\traj)$. Then, we can extract
$\Delta \f_\class$ from the class average of the exponential entropy
difference:
\begin{align*}
\avgatclass{e^{-\Entlike}} & = \avgatclass{e^{-\beta(W-\Delta \f)}}
  \nonumber \\
  & = e^{\beta \Delta \f_\class}\avgatclass{e^{-\beta W}}
  ~.
\end{align*}
Then, the ECFT Eq. (\ref{eq:exptcft})
gives:
\begin{align}
  \label{eq:const_f_exp}
  \Delta \f_\class
  = -\beta^{-1} \ln \langle e^{-\beta W} \rangle_\class
+ \beta^{-1} \ln \frac{\revP(\class)}{\forP(\class)}
  ~.
\end{align}
This equality relates free-energy differences to statistics on works and class
probabilities. Jarzynski's Equality Eq. (\ref{eq:jarzynski}) is the special
case where $\class = \trajs$, $\distinit=\eqdist_0$, and
$\distinittwo=\eqdist_\tau$.

\subsection{Metastable Process Work Bounds}
\label{sec:metastable_process_wb}

Having established Eq. (\ref{eq:const_f_exp}), we now demonstrate its
application to bounding the ensemble average work of a
particular type of process that we call a \emph{metastable process}.

As a motivating example, consider an information-storing biomolecule
whose configurational space is too complex to fully model but which has
a coarser description of state that is robust to thermal noise, like
the various functional configurations of a protein or RNA molecule. With the
results here, one can obtain refined bounds for the work production in
altering the occupancy of these coarsened states along with free energies
associated with these states, without
knowing the exact details of the underlying Hamiltonian. A similar analysis
to ours was done to obtain the change in free energy of RNA through stretching
\cite{Liph02a, Juni09a}.
However, our procedure comes with the added possibility of treating
distributions over multiple possible coarsened states.

\subsubsection{Metastable Processes}

Consider an energetic landscape $\Eat t$ at time $t$ that is partitioned into
regions---\emph{metastable regions}---each separated by high energy barriers.
For a system contained in any such region, the barriers severely limit the
chance of escape over long timescales. Each metastable region therefore
represents an information-storing mesostate, or \emph{memory state}, that
robustly constrains the system. Also, for any system state distribution that has
support over exactly one metastable region $m$, the system will relax to a
stable distribution $l_t^m$ over the region much faster than the timescale of
escape if the energetic landscape is left unperturbed.  While not the true,
global equilibrium distribution $\eqat t$, which generally has support over all
metastable regions, we call $l_t^m$ the local equilibrium distribution for $m$.

Now, prepare the system in arbitrary distributions over all of phase space and
then allow the system to locally equilibrate in $E_t$. We call the system state
distribution $\dist$ obtained after local equilibration a \emph{metastable
distribution}. Within each metastable region $m$, $\dist$ must match $l_t^m$ up
to normalization and, therefore, the equilibrium distribution in that region.
Thus, we have:
\begin{align*}
  \dist(\mst) = \dist(\metregion(\mst))
  \frac{\eqat{t}(\mst)}{\eqat{t}(\metregion(\mst))}
  ~,
\end{align*}
where $\metregion(\mst)$ is the metastable region for microstate $\mst$,
the probability of the system being in the metastable region $m$ is defined
as $\dist(m) \equiv \int_m d\mst \, \dist(\mst)$, and
$\eqat{t}(m)\equiv\int_m d\mst \, \eqat{t}(\mst)$ is the
equilibrium probability of being in the region $m$.

A metastable process is then a forward process where (i) the initial and final
energetic landscapes $\Eat0$ and $\Eat\tau$ can each be partitioned into
metastable regions and (ii) the initial distribution $\distinit$ is metastable
over $\Eat0$.

\subsubsection{Metastable Free Energies}

For a metastable distribution $\dist$, all system states in a metastable region
$\metregion$ have the same free energy. That is, for $\mst \in \metregion$:
\begin{align*}
  f(\mst; \dist, \Eat{t})
  &= \Eat{t}(\mst) + \beta^{-1}\ln\dist(\mst) \\
  &= \Eat{t}(\mst) + \beta^{-1}\ln\dist(\metregion(\mst))
  \frac{\eqat{t}(\mst)}{\eqat{t}(\metregion(\mst))} \\
  &= \Eat{t}(\mst) + \beta^{-1}\ln \frac{\dist(\metregion(\mst))}
  {\eqat{t}(\metregion(\mst))}e^{-\beta (\Eat{t}(\mst)-\Feq_t)} \\
  &= \beta^{-1}\ln \frac{\dist(\metregion(\mst))}{\eqat{t}(\metregion(\mst))}
  + \Feq_t
~.
\end{align*}
Thus, the free energy for a locally equilibrated distribution over a
metastable region is the free energy of any state in the region.
We call such a free energy a \emph{metastable free energy}.

This further simplifies if we identify the \emph{memory-state free energy}:
\begin{align}
  F^\text{mem}_t(\metregion) \equiv F^\text{eq}_t
  - \beta^{-1} \ln \eqat{t}(\metregion)
  ~,
\end{align}
as the fixed contribution of a particular memory state to the free energy,
regardless of the metastable distribution $\dist$. The free energy is thus
the free energy of the memory state plus the surprisal of that memory state:
\begin{align*}
  f(\mst; \dist, \Eat{t})
  = F^\text{mem}_t(\metregion(\mst)) + \beta^{-1}\ln\dist(\metregion(\mst))
  ~.
\end{align*}  
When averaged over all metastable regions with distribution $\kappa$, this
returns a familiar expression \cite{Parr15a} for average nonequilibrium free
energy:
\begin{align*}
  \avgat{f}{\mathcal{Z}}(\kappa,E_t)
  = \sum_{m} \kappa (m) F^\text{mem}_t(m)-\beta^{-1}H_M(\kappa)
  ~,
\end{align*}
where $H_M(\kappa) \equiv -\sum_m \kappa(m) \ln \kappa(m)$ is the Shannon
entropy of $\kappa$ over memory states---the average amount of information they
store.

This decomposition offers an entr\'ee to the problem of evaluating work
production of a process whose control protocol $\vec{\lambda}$ must start
in a particular initial configuration $\lambda(0)$ and end in a particular
final configuration $\lambda(\tau)$. If their respective energetic landscapes
$E_0$ and $E_\tau$ are not well understood, we can still extract bounds on work
production using the TCFT.

We wish to derive an ensemble average free-energy difference for such a
metastable process so that we can bound the work invested in the forward
experiment that exploits the simplicity of metastable free energies.
Of course, different choices of the second privileged distributions
$\distinittwo$ result in different free-energy differences, but we will
consider the metastable distribution that corresponds to the final-time
distribution of the metastable process.
That is, we choose $\distinittwo$ to be the distribution of the system if,
holding the energetic landscape fixed at $\Eat\tau$ at the end of the
protocol, the system locally-equilibrated after the end of the
forward process:
\begin{align*}
  \distinittwo(\mst) = \distfinal(\metregion(\mst))
  \frac{\eqat\tau(\mst)}{\eqat\tau(\metregion(\mst))} 
  ~.
\end{align*}
We say that $\distinittwo$ is then the locally-equilibrated distribution of
$\distfinal$.
The resulting free-energy difference $\Delta\f$ is then called the
\emph{metastable free energy change} for the metastable process:
\begin{align*}
  \Delta\f(\traj)
  &=F_\tau^\text{mem}(\metregion'(\trajat0))
  -F_0^\text{mem}(\metregion(\trajat\tau)) \\
  &\quad+\beta^{-1} \ln\frac{\distinittwo(\metregion'(\trajat\tau))}
  {\distinit(\metregion(\trajat0))}
  ~,
\end{align*}
where $\metregion(\mst)$ and $\metregion'(\mst)$ are the memory states
containing $\mst$ in $\Eat0$ and $\Eat\tau$, respectively.
Using Eq. (\ref{eq:wf_bound}), we then have:
\begin{align}
\label{eq:MetaBound}
\avgatens{W} & \geq \avgatens{\Delta f}
\\ &= \avgatens{\Delta F^\text{mem}} - \beta^{-1}\Delta H_M \nonumber
~.
\end{align}
The first term captures the free energy contribution of each
state and is specific to the particular physical instantiation of the memory.
The second term is characteristic of how the system's distribution
over metastable regions transformed.
If the memory states all had the same free energy then the first term would be
zero, giving Landauer's bound.
Generally, Eq. (\ref{eq:MetaBound}) falls short of the free-energy change
bound on the average work, Eq. (\ref{eq:wf_strictest_bound}), because
the actual final-time free energy, that of $\distfinal$, is generally higher
than the free energy of the locally-equilibrated $\distinittwo$.

\subsubsection{Obtaining Memory-State Free Energies}

\newcommand{\pros}{\mathsf{P}}

Consider a system and a pair of initial and final protocol configurations
$\lambda(0)$ and $\lambda(\tau)$, respectively.  If each of these
configurations contains metastable regions, capable of storing useful
information, it is worthwhile considering the family of thermodynamic
experiments that execute computations, stochastically mapping between different
metastable regions of these end-points.  As we will show, any experiment that
begins in a metastable distribution with the boundary conditions above obeys
strong bounds on work production related to the metastable free energy.

This is useful since we can choose one or a small number of such processes to
study in detail to obtain the memory state free energies and then, by Eq.
(\ref{eq:MetaBound}), the average work for any computation implemented between
these control points is simply determined by the initial and final memory state
distributions.

Consider an initial metastable region $\metregion$ and final metastable region
$\metregion'$ and the associated trajectory class that connects them:
\begin{align*}
  \class_{\metregion, \metregion'}
  = \{\traj | \trajat0 \in \metregion, \trajat\tau \in \metregion'\}
  ~.
\end{align*}
All trajectories within one class must all have the same free energy difference
if we choose the privileged distributions $\distinit$ and $\distinittwo$ of our
forward and reverse experiment to be metastable. Practically, this can be
experimentally implemented by allowing the system to relax to local metastable
equilibria before executing the control protocol. If the metastable regions are
appropriately chosen, this process can be much faster than relaxation to global
equilibrium.

As a result, we can express the free-energy difference for this trajectory
class in terms of the input and output memory states by considering $\traj \in
\class_{\metregion,\metregion'}$:
\begin{align*}
  \Delta \f_{\class_{\metregion, \metregion'}}
  &= \f(\trajat\tau;\distinittwo,E_\tau) - f(\trajat0;\distinit,E_0) \\
  &= F^\text{mem}_\tau(\metregion')-F^\text{mem}_0(\metregion)
  + \beta^{-1} \ln \frac{\disttwo (\metregion')}{\distinit (\metregion)}
~.
\end{align*}
In the special case where $\distinit(m)=\disttwo(m')$, the
change in memory state free energy is equal to the free-energy difference.

Regardless of the metastable privileged distributions used, we
recover memory state free-energy differences via Eq. (\ref{eq:const_f_exp}),
using the work production probabilities and memory state probabilities:
\begin{align*}
  &F^\text{mem}_\tau(\metregion')-F^\text{mem}_0(\metregion) \\
  &\quad= -\beta^{-1} \ln
  \avgat{e^{-\beta W}}{\class_{\metregion,\metregion'}}
+ \beta^{-1} \ln\frac
  {\revP(\class_{\metregion,\metregion'})\distinit(\metregion)}
  {\forP(\class_{\metregion,\metregion'})\distinittwo(\metregion')}
  ~.
\end{align*}

If the protocol is cyclical, such that $E_\tau=E_0$, then we can
fully obtain all memory state free-energy differences with $|\mathcal{M}|-1$
trajectory classes, where $\mathcal{M}$ is the set of metastable regions of the
initial-final energetic landscape. Otherwise, the memory state free-energies
can be determined by considering $|\mathcal{M}_1| + |\mathcal{M}_2|-1$
trajectory classes, where $\mathcal{M}_1$ and $\mathcal{M}_2$ are the sets of
metastable regions of the initial and final landscapes.

With the free energy landscapes determined, it becomes straightforward to set
strong bounds not only on the process that resulted from the original
experiment with $\vec{\lambda}$ and $\distinit$, but on \emph{any} experiment
that has the same initial and final protocol configurations and begins in a
metastable distribution. Suppose the initial memory state distribution of the
process is $q$. Let the resultant final memory state distribution of the
process be $q'$:
\begin{align*}
  q'(\metregion') =\sum_\metregion q(\metregion)
  p_{\metregion \rightarrow \metregion'}
  ~,
\end{align*}
where $p_{\metregion\rightarrow\metregion'}$ is the probability of the system
ending in $\metregion'$ given that it started in $\metregion$. Then, by Eq. (\ref{eq:MetaBound}):
\begin{align*}
\avgatens{W}
  \geq& \sum_{\metregion'}q'(\metregion')F^\text{mem}_\tau(\metregion')
  -\sum_{\metregion}q(\metregion)F^\text{mem}_0(\metregion) \\
  & - \beta^{-1}(H_M(q')-H_M(q))
  ~.
\end{align*}

With this section's results, it is possible to obtain strong work bounds on
computations even when the memory-state free energies are unequal or unknown at
the outset. This significantly strengthens the Second Law as applied to the
thermodynamics of computation.

\section{Statistical Freedom from the Tyranny of the Rare}
\label{sec:tyranny}

When estimating statistical quantities from data, rare events can dominate
sample averages \cite{Jarz06a, Asba17a}. This can be particularly problematic
when the events are associated with large resources. Consider the following
case in point. By empirically estimating the exponential average work
$\avgat{e^{-\beta W}}{\trajs}$ for a thermodynamic transformation, one can
estimate the equilibrium free energy difference $\Delta \Feq$ via Eq.
(\ref{eq:jarzynski})'s Jarzynski's Equality. However, this can require thorough
sampling of very rare events \cite{Jarz06a}. The \exptcftabbr\, of Eq.
(\ref{eq:exptcft}) can aid in solving this statistical challenge by removing
consideration of these rare but work-dominant trajectories.

Specifically, if the privileged starting distributions of both the
forward and reverse experiments are in equilibrium, then the free energy
difference $\Delta f_C$ of a class $C$ (shown in Eq. (\ref{eq:const_f_exp})) is
the change in free energy $\Delta F^\text{eq}$, regardless of the chosen
class. However, given a set of $N$ forward and reverse experiments and
associated work data for the forward experiment
$\vec{W}=\{W_1,W_2, \cdots W_N \}$, statistical fluctuations in the data
lead to fluctuations in trajectory class probability estimates $\tilde{P}(C)$
and $\tilde{R}(C)$, where the tilde indicates an estimate. This results in
statistical fluctuations in free-energy difference estimates that depend on
the trajectory class:
\begin{align}
\Delta \tilde{f}_C \equiv - \beta^{-1} \ln \left( \frac{\sum_{i=1}^N
 \delta_{W_i \in C} e^{-\beta W_i}}{\sum_{i=1}^N \delta_{W_i \in C} } \right)
 + \beta^{-1}\ln \frac{\tilde{R}(C)}{\tilde{P}(C)},
\end{align}
where $\delta_{W_i \in C}$ returns $1$ if the $i$th work value is realized
within the trajectory class $C$ and $0$ otherwise. If we had perfect
statistics, these estimates would all be the actual change in free energy
$\Delta F^\text{eq}$. However, as the next section shows (and as App.
\ref{app:Free Energy Estimate Distribution} further explains), we can find
better estimators of the free energy change by choosing more probable
trajectory classes.

\subsection{Tyranny of the Rare}

Consider the following example 1D system.  It is in contact with a thermal
environment at inverse temperature $\beta$ but otherwise obeys classical
mechanics under a time-evolving potential energy landscape. There exist two
regions in state space, $A$ and $B$, each with potential energy that is
constant over their regions. Arbitrarily high barriers separate and surround
the regions so that all particles in a given region stay there. These two
potential-energy wells start at energies:
\begin{align*}
  E_0(A) &= -\beta^{-1} \ln(1-\epsilon)
\end{align*}
and:
\begin{align*}
  E_0(B) &= -\beta^{-1} \ln(\epsilon)
  ~,
\end{align*}
respectively,
where $0 < \epsilon \ll 1$ and the energy everywhere else is arbitrarily
large.

Start the system in equilibrium over the two wells so that the probabilities
of starting in the wells are:
\begin{align*}
  P(A) &= 1 - \epsilon
\end{align*}
and:
\begin{align*}
  P(B) &= \epsilon
  ~.
\end{align*}
Now, raise well A and lower well B to end at
$E_\tau(A)=E_\tau(B)=\beta^{-1} \ln 2$ energy. Consider a trajectory class
for trajectories that are wholly in the $A$ well during the process and
similarly a class for the $B$ well.  We refer to the classes synonymously
with their associated wells.

The work invested for either class is simply the change in energy of the
corresponding well since the energy barrier is high enough that system
states do not cross between wells during the control protocol:
\begin{align*}
  W_A & = \beta^{-1}\ln (2 (1-\epsilon)) \\
  W_B  & = \beta^{-1}\ln (2 \epsilon)
  ~.
\end{align*}
The resulting integral fluctuation theorem yields:
\begin{align*}
\avgat{e^{-\beta W}}{\trajs}
  &= \forP(A) \avgat{e^{-\beta W}}{A}
    + \forP(B) \avgat{e^{-\beta W}}{B} ~,
\end{align*}
where:
\begin{align*}
\forP(A)\avgat{e^{-\beta W}}{A}
  &= (1-\epsilon) \frac{1}{2 (1-\epsilon)} \nonumber \\
  & = \frac{1}{2}
  ~,
\end{align*}
and:
\begin{align*}
\forP(B) \avgat{e^{-\beta W}}{B}
  &= \epsilon \frac{1}{2\epsilon} \nonumber \\
  & = \frac{1}{2}
  ~.
\end{align*}
So, the total exponential average work is:
\begin{align*}
  \avgat{e^{-\beta W}}{\trajs}
= 1
  ~,
\end{align*}
and, thus, by Eq. (\ref{eq:jarzynski}) the equilibrium free energy change
vanishes.

\begin{figure*}
\includegraphics[width=2\columnwidth]{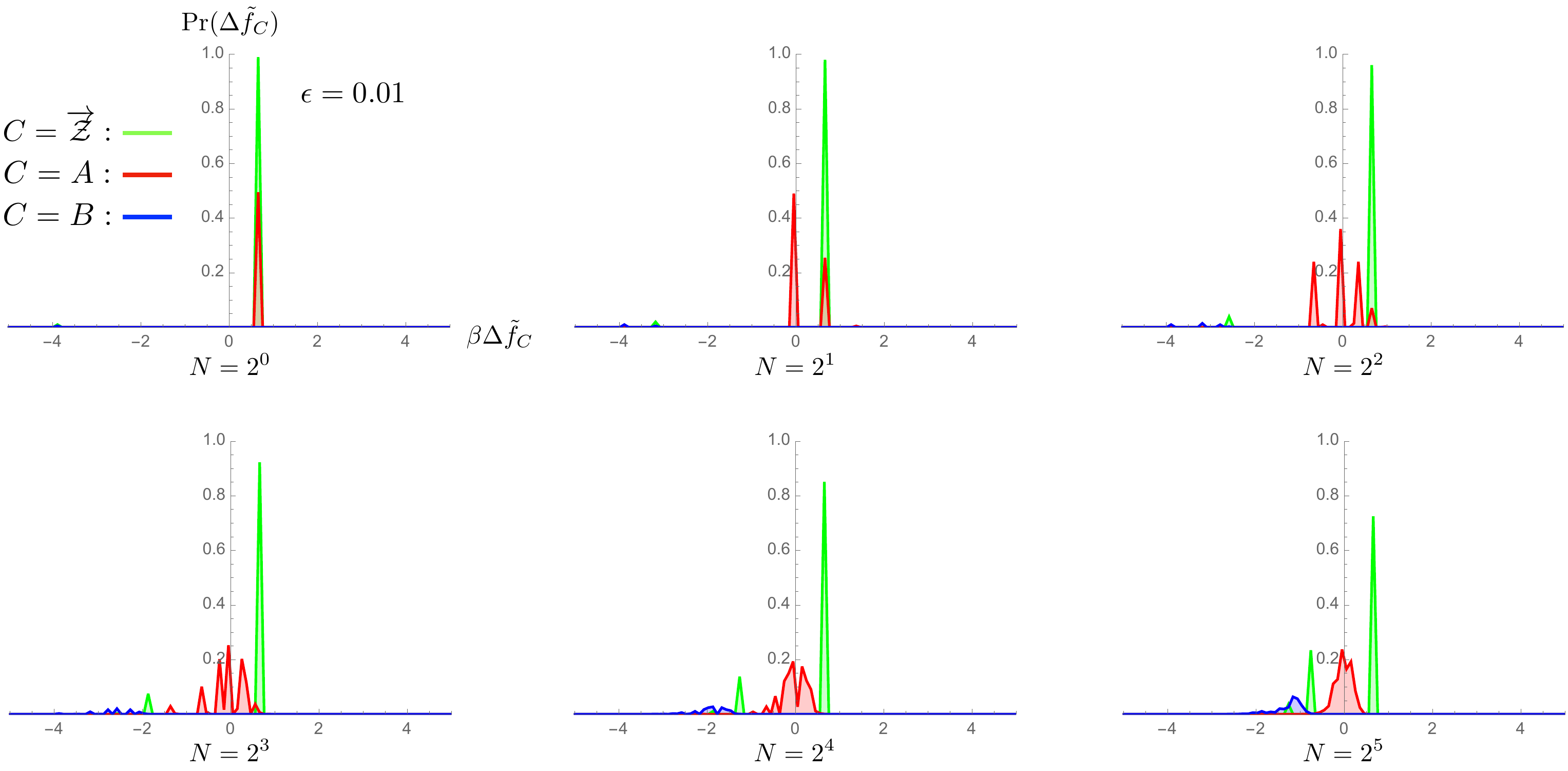}
\caption{Tyranny of the rare: Distribution of free energy estimates
	$\Delta \tilde{f}_C$ arising from $N$ forward and $N$ reverse experiments
	with $\epsilon=0.01$ depends sensitively on the trajectory class $C$.
	Above, plots of free energy estimates for the trajectory class that
	includes all paths (green: $C= \text{all}$), the trajectory class that
	starts and ends in $A$ (red: $C=A$), and the trajectory class that starts
	and ends in $B$ (blue: $C=B$). Note that we do not plot divergent free
	energy estimates, for which we estimate $\tilde{P}(C)=0$ or
	$\tilde{R}(C)=0$. Both trajectory classes $A$ and $B$ can yield divergent
	estimates, but $A$ often provides better estimates than $C=\text{All}$.
	}
\label{fig:FreeEnergyHistogram} 
\end{figure*}

In this situation, the probabilities of the two classes are highly uneven with
class $B$ having only probability $\epsilon$. However, $B$ accounts for $1/2$
of the total exponential average work. This means that an accurate statistical
estimate of the change in equilibrium free energy via sampling such a process
and using Eq. (\ref{eq:jarzynski}) is highly dependent on rare events.
Specifically, even though the true free energy change is $0$, it would likely
be estimated as $\Delta \tilde{F}^\text{eq}=\beta^{-1}\ln 2$ with small enough
$\epsilon$, as can be seen by the green histogram in Fig.
\ref{fig:FreeEnergyHistogram}. It converges to the true value only with very
large samples of the rare class $B$. Thus, the variance of estimated values for
the change in equilibrium free energy is large for finitely sampled
experiments. Typically, estimates will be misleading.

\subsection{Circumventing Tyranny}

The TCFT solves this problem using appropriate trajectory classes. In
principle, we may consider a process with arbitrary privileged distributions
for both the forward and reverse processes. Thus, we can estimate arbitrary
free-energy differences for a process, as long as we restrict consideration to
a trajectory class $\class$ of constant free-energy difference, as described in
Sec. \ref{sec:const_f_theory}. However, constant free-energy differences are
automatically guaranteed for any class when we choose equilibrium privileged
distributions for both the forward and reverse processes.  The privileged
distribution for the forward process described above was indeed equilibrium and
we choose an equilibrium distribution for the corresponding reverse process as
well.

Now, focus on Eq. (\ref{eq:const_f_exp}), moving away from Jarzynski's Equality
in Eq.  (\ref{eq:jarzynski}). This expands the required estimators from simply
the exponential average work to also include the forward and reverse process
probabilities of class $\class$. Moreover, the exponential average work is now
conditioned on $\class$. Thus, the estimator is a function of sampled data that
comes from class $\class$ in both the forward and reverse experiments. However,
we need $\class$ to be such that the class average of the exponential work, the
forward probability of the class, and the reverse probability of the class are
statistically easy to estimate.

This is the case when sampling trajectories from
class $A$ in the example above. First, note that $A$ has very high probability $1 - \epsilon$, so
estimating its log-probability from data is statistically easy. Second, its class
average exponential work is also easily estimated since the class itself is
highly likely and its work distribution is narrow.

This leaves estimating the reverse class probability. In this, we choose our
second privileged distribution to be the equilibrium distribution for the
final-time energy landscape and so solve for the equilibrium free-energy
change, as desired. This, then, fully specifies the reverse process. Since the
true equilibrium free-energy change vanishes, Eq. (\ref{eq:const_f_exp}) says
that the reverse-process probability of class $A$ is $\revP(A) = 1/2$. In this
way, since $A$ is likely in the reverse process, it too is easily estimated.

To verify that this reasoning is sound, consider estimates obtained from
various numbers $N$ of forward and reverse process trajectories for the three
trajectory classes $\trajs$, $A$, and $B$. (See App. \ref{app:Free Energy
Estimate Distribution} for an explanation of the base calculations.) Figure
\ref{fig:FreeEnergyHistogram} shows that the distribution of free energy
estimates when using all trajectories $C=\trajs$ is heavily weighted towards
$\Delta \tilde{f}_{\trajs}=\beta^{-1} \ln 2$ for $\epsilon=0.01$ and a small
data set of work values. However, if we restrict to the trajectory class that
starts and ends in the most likely state $A$, then we see that free energy
estimates are more closely centered around the correct value of $\Delta
F^\text{eq}=0$. This suggests that restricting to high probability regions of
the work distribution improves free energy estimates.

\begin{figure*}
\includegraphics[width=2.05\columnwidth]{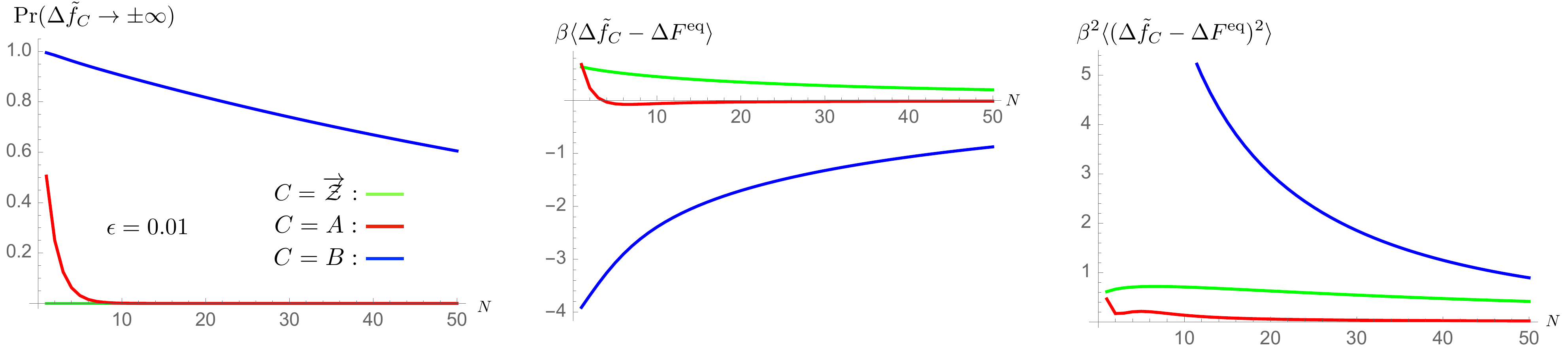}
\caption{Degree to which energy estimates $\Delta \tilde{f}_C$ diverge from the
	actual change in free energy $\Delta F^\text{eq}$ depends on the trajectory
	class: For $\epsilon=0.01$, the probability of an infinitely
	divergent free energy estimate $\Pr(\Delta \tilde{f}_C \rightarrow \pm
	\infty)$ is significant for the trajectory class $B$, nonzero and swiftly
	decreasing for $A$ with larger data, and zero for the set of all
	trajectories. Despite this advantage in using all trajectories to estimate
	free energy, both the average difference $\langle \Delta
	\tilde{f}_C-\Delta F^\text{eq}\rangle$ and the mean squared deviation
	$\langle (\Delta \tilde{f}_C-\Delta F^\text{eq})^2\rangle$ are minimized by
	the class $C=A$ by excluding the rare event $B$. These plots come from
	excluding infinitely divergent estimates, which represent an exponentially
	small likelihood for $A$ as $N$ increases. Looking at a reduced trajectory
	subspace $A$ gives consistently improved estimates for small amounts of
	data.
	}
\label{fig:FreeEnergyConvergence} 
\end{figure*}

Figure \ref{fig:FreeEnergyConvergence} further quantifies this advantage by
plotting the average difference with the correct free energy $\langle \Delta
\tilde{f}_C-\Delta F^\text{eq} \rangle$ and mean square deviation $\langle
(\Delta \tilde{f}_C-\Delta F^\text{eq})^2 \rangle$. We see a marked advantage
to our restricted trajectory class $C=A$ over the full set of trajectories in
both cases.

By contrast, restricting to the rare event $C=B$, Fig.
\ref{fig:FreeEnergyHistogram} shows that the majority of the free energy
estimates are divergent for a small data set of work values. However, even when
considering only the nondivergent estimates, Fig.
\ref{fig:FreeEnergyConvergence} shows that the rare trajectory class $C=B$
leads to the worst free energy estimates.

The resulting estimate of the free-energy change via the TCFT is indeed much
more statistically robust and parsimonious. Reference \cite{Asba17a} gives
another analysis focusing on improvements in reducing the bias.
The approach detailed above shows how we can reduce the bias as well
reduce the variance. 

\section{Related Results}
\label{sec:related_results}

We now turn to the burgeoning collection of previously established fluctuation
theorems. As noted, many existing fluctuation theorems are special cases of
the TCFT.

For most of the results and setups considered in the following, the
trajectory classes constructed can have zero probability. Of course, the TCFT
is then not to be used directly. Instead, the methods of Sec.
\ref{sec:zero_probs} can be used to find either approximate trajectory classes
or a limiting procedure that will then yield the exact result in question. To
avoid redundancy, we simply assume a limiting procedure when the trajectory
class has zero probability.

\subsection{Measurement and Feedback for Free Energy Estimation}

Reference \cite{Asba17a} focuses on the problem of estimating free energy
differences in the face of rare yet resource-dominant events. This was the
subject of Sec. \ref{sec:tyranny}. By taking measurements of the system state
at some number of times and rejecting or accepting data samples based on those
measurement results, statistical estimation of free energies can be improved
with the use of their Eq. (9).

Specifically, suppose that $A=\{A_1,\ldots,A_N\}$ is a set of regions of
system state space. For times $T'=\{t_1,\ldots,t_N\}$ during the
protocol, we measure whether the system occupies region $A_i$ at time
$t_i$ for each $i$. Let the class $\class$ be trajectories that occupy
$A_i$ at time $t_i$ for each $i$. Applying Eq. (\ref{eq:exptcft})'s ECFT to
$\class$, we derive their Eq. (9). They show that an estimator of the
free-energy difference over a process can have a smaller bias when using their
Eq. (9) and an appropriate set of measurement choices compared to using the
Jarzynski Equality.

Since $\class$ incorporates an arbitrary number of arbitrary system state
measurements, their Eq. (9) is a rather broad fluctuation theorem. The key
difference between Eq. (\ref{eq:exptcft}) and their Eq. (9) is that Eq.
(\ref{eq:exptcft}) allow (i) an infinite number of specifications on the
trajectories and (ii) more general types of trajectory specifications (e.g.,
work values) than instantaneous descriptions of the system state.

\subsection{Phase Space Perspective on Dissipated Work}

Reference \cite{Kawa07a} achieves several important results that can be
understood as special cases of the TCFT. They consider a system that only
interacts with a control device during the forward and reverse experiments.
Note that this is a special case of our assumptions where the system has
negligible or no interaction with the thermal environment. Also note that, as
their examples illustrate, this still allows a system to be composed of two
subsystems with one acting as a thermal environment for the other. In these
cases, the TCFT can be applied to either the entire system or the latter
subsystem. We apply it to the entire system to follow their core development.

Set the state space $\msts$ to be the microstates of the system and the modeled
times $T$ to be $[0,\tau]$. This ensures a deterministic, Hamiltonian evolution
of the system, as they describe.
Set $\distinit$ and $\distinittwo$ to the initial and final Boltzmann
distributions at the same inverse temperature $\beta$. So, the resulting
free-energy differences of all trajectories are equal to the equilibrium
free-energy difference $\Delta\Feq$. The dissipated work
$\avgatens{W}-\Delta\Feq$ is then the average heat that would be dissipated
from the system to a thermal environment at inverse temperature $\beta$ if the
system was equilibrated with the environment after the forward protocol while
held at energetic landscape $\Eat{\tau}$.

Then consider a partition $\{\chi_1,\ldots\chi_K\}$ of $\msts$ and a time $t$.
For each $j$ in $\{1,\ldots,K\}$, let $\class_j$ be the set of system state
trajectories that occupy $\chi_j$ at time $t$. Then let $\rho_j$ be the
probability the system is in $\chi_j$ at time $t$ in the forward experiment.
Then:
\begin{align*}
  \rho_j=\forP(\class_j)
  ~.
\end{align*}
Set $\widetilde\chi_j=\rev{\chi_j}=\{\rev\mst\,|\,\mst\in\chi_j\}$ and let
$\widetilde\rho_j$ be the probability the system is in $\widetilde\chi_j$ at
time $\tau-t$ in the reverse experiment. (They denote this time $t$, keeping
all references of time to be relative to the forward experiment.) Then:
\begin{align*}
  \widetilde \rho_j & = \physrevP(\rev{\class_j}) \\
  & = \revP({\class_j})
  ~.
\end{align*}
Applying Eq. (\ref{eq:exptcft})'s ECFT to $\class_j$ yields Ref.
\cite{Kawa07a}'s Eq. (6). Eq. (\ref{eq:tcsl})'s TCSL applied to $\class_j$
yields their Eq. (7). And, Eq. (\ref{eq:tpsl})'s TPSL applied to the partition
of trajectories $Q=\{\class_1,\ldots,\class_K\}$ yields their Eq. (8).

These results, especially their Eq. (8), were used to provide concise
expressions involving work and free energies for simple but instructive
processes, as well as to derive Landauer's bound.

\subsection{Work Dissipation as the Distance from Equilibrium}

Reference \cite{Vaik09a} obtains an inequality between the dissipated work up
to any time $t$ during a protocol and how far the system's state density at
time $t$ must be from equilibrium. They assume that the system dynamics are
Markovian and that the system equilibrates, at any time $t$, towards
the Boltzmann distribution for $E_t$ and $\beta$, if the protocol is suddenly
interrupted at time $t$ and the system is held under energetic landscape
$\Eat{t}$. These assumptions are met in our model if the thermal environment
has such a high relaxation rate that it is effectively memoryless as far as the
influence on the system is concerned. They also assume that $\distinit$ and
$\distinittwo$ are the initial and final Boltzmann distributions. Let $\prot$
be the forward protocol, which runs from time $0$ to $\tau$. Let $\msts$ be the
system microstates and $T=[0,\tau]$.

We derive their results from the TCFT by considering a separate protocol
$\prot^t$ for any time $t$ in $[0,\tau]$. $\prot^t$ runs from time $0$ to time
$2t$. For $t'\leq t$, $\prot^t(t')=\prot(t')$. For $t'>t$,
$\prot^t(t')=\prot(t)$. Thus $\prot^t$ follows $\prot$ until time $t$, at
which point $\prot^t$ holds fixed until it ends at $2t$. The forward process
privileged distribution $\distinit^t$ for the protocol $\prot^t$ is simply set
to the Boltzmann distribution $\distinit$. At time $t$, denote the probability
of being in state $\mst$ as $\rho(\mst,t)$, which must be shared between both
protocols $\prot$ and $\prot^t$ since the two protocols do not differ until
time $t$.

The reverse process privileged distribution for $\prot^t$ is set to
be Boltzmann with respect to $\Eat{2t}$. Since the protocol $\prot^t$
is fixed between times $t$ and $2t$, this is also the physical-reverse state
distribution at all times between $0$ and $t$ during the physical-reverse
process. In particular, the state distribution for the physical-reverse
process of protocol $\prot^t$ at time $t$ is:
\begin{align*}
  \rho^\text{eq}(\mst,\prot(t))=e^{-\beta(\Eat{t}(\mst)-\Feq_t)}
  ~.
\end{align*}

Let $\class^t_\mst$ be the set of trajectories that occupy microstate $\mst$
at time $t$.  Then:
\begin{align*}
  \rho(\mst,t)=\forP(\class^t_\mst)
  ~,
\end{align*}
where $\forP$ refers to the forward process of protocol $\prot^t$.
And:
\begin{align*}
  \rho^\text{eq}(\mst,\prot(t))
  &=\physrevP(\class^t_\mst) \\
  &=\revP(\class^t_\mst)
  ~,
\end{align*}
where $\physrevP$ and $\revP$ refer to the reverse process (physical and
formal representations, respectively) of protocol $\prot^t$.

Let $W(t)$ denote the work conducted up to time $t$. The exponential average work up to time $t$ conditioned on the system occupying state $\mst$ at time
$t$ is $\avgat{e^{-\beta W(t)}}{\mst,t}$, during either protocol $\prot$ or
$\prot^t$. Since no additional work is conducted under the protocol $\prot^t$
after time $t$, this quantity is then:
\begin{align*}
  \avgat{e^{-\beta W(t)}}{\mst,t}
  =\avgat{e^{-\beta W}}{\class^t_\mst}
  ~,
\end{align*}
where the second average is taken over the forward process for protocol
$\prot^t$.

Then, applied to the protocol $\prot^t$ and using trajectory class
$\class^t_\mst$ and the above equalities, Eq. (\ref{eq:exptcft})'s ECFT yields
Ref. \cite{Vaik09a}'s Eq. (6). Equation (\ref{eq:tcsl})'s TCSL yields their Eq.
(8) and Eq. (\ref{eq:tpsl})'s TPSL yields their Eqs. (2) and (9).

\subsection{Work and State Fluctuation Theorem}

Consider a process where $\distinit$ and $\distinittwo$ are the initial and
final Boltzmann distributions.
Reference \cite{Mara08a} establishes a fluctuation theorem that relates a work
value, the equilibrium free-energy difference, and forward and reverse process
probabilities for obtaining the work value and particular values for two
functions of state. The two functions of state are evaluated at opposite ends
of the trajectory. In their notation, this is written as:
\begin{align*}
  \forP_\text{F}(\widetilde W,a\rightarrow b)e^{-\beta \widetilde W}
  =\forP_\text{R}(-\widetilde W,b^*\rightarrow a^*)e^{-\beta\Delta\Feq}
  ~,
\end{align*}
which is their Eq. (1). (Except that we use $\widetilde W$ for a specific work
value.) Here, $a$ and $b$ are some output values of the two respective
functions of state and $a^*$ and $b^*$ are the output values of the time
reverse of system states that output $a$ and $b$. $\forP_\text{F}$ and
$\forP_\text{R}$ are the forward and reverse processes.

Let $\class$ be all trajectories (i) that obtain a work value $W$, (ii)
whose state evaluates the first state function to $a$, and (iii) whose end
state evaluates the second state function to $b$.  Then applying the
ECFT Eq. (\ref{eq:exptcft}) to $\class$ gives their Eq. (1).

The equality was used to efficiently estimate the conformational free energy
change of a simulated alanine dipeptide.

\subsection{Landauer's Bound on Erasure Dissipation}

Finally, Ref. \cite{Beru13a} considers the process of erasing information
implemented using a Brownian silica bead trapped in an optical tweezer. The
laser initially induces an effective double-well potential that is symmetric
across the center. They call the two wells $0$ and $1$. Manipulating the laser
and moving the platform containing the bead, a protocol is realized that, while
ending with the effective potential in the initial form, shifts the bead to the
$0$ well with high probability. This was obtained with a variety of specific
initial conditions and protocol variations. Since the bead always starts in
either well with $50\%$ probability, this then demonstrates erasing one bit of
information via a variety of protocols.

They also demonstrate that the average work required to conduct these
protocols was always near $\beta^{-1}\ln2$, verifying Landauer's Bound.  To
explain why Landauer's bound should hold, they utilize Ref. \cite{Vaik09a}'s
Eq. (6) to produce the following:
\begin{align}
  \label{eq:berut0}
  \avgat{e^{-\beta W}}{\rightarrow 0}=\frac{1/2}{P_S}
  ~,
\end{align}
and:
\begin{align}
  \label{eq:berut1}
  \avgat{e^{-\beta W}}{\rightarrow 1}=\frac{1/2}{P_S}
  ~,
\end{align}
where $P_S$ is the probability of a trajectory successfully ending in the $0$
state, and $\rightarrow 0$ ($\rightarrow 1$) denotes an average conditioned on
ending in the $0$ ($1$) state. In particular, applying Jensen's inequality to
Eq. (\ref{eq:berut0}) yields:
\begin{align*}
  \avgat{W}{\rightarrow 0}\geq\beta^{-1}(\ln2 + \ln P_S)
  ~,
\end{align*}
which is a generalization of Landauer's Bound for imperfect erasure.

We deduced Ref. \cite{Vaik09a}'s Eq. (6) from the TCFT already, but Eqs.
(\ref{eq:berut0}) and (\ref{eq:berut1}) can be obtained from the TCFT directly
and quickly. Since the bead starts off equilibrated over each well and has
equal probability to start in either, the initial distribution $\distinit$ is
equilibrium. Choose $\distinittwo$ to also be equilibrium. Then the free energy
difference is $\Delta\Feq$, which must be zero since the effective potential
ends as it begins. Then let $\class_0$ be all trajectories that end in $0$ and
$\class_1$ be all that end in $1$. Applying the ECFT Eq. (\ref{eq:exptcft})
first to $\class_0$ and then to $\class_1$ yields Eqs. (\ref{eq:berut0}) and
(\ref{eq:berut1}), respectively.

\section{Discussion}
\label{sec:discussion}

The preceding provided guidance when selecting appropriate trajectory
classes for a given process of interest. To achieve a much stronger bound on
entropy difference than the traditional ensemble-average Second Law, Sec.
\ref{sec:tpsl} proposed choosing a partition of all trajectories into classes
that resulted in narrow entropy difference
distributions for each class. And, to avoid dominating rare events when
calculating free-energy differences (Sec. \ref{sec:tyranny}), we recommended
classes that are common in both the forward
and reverse processes and that have narrow entropy-difference distributions.
That said, the most effective classes for these tasks appear to be specific
to the particular processes of interest. Developing procedures to identify
these classes for arbitrary processes remain an open problem.

What does this look like in practice? Reference \cite{Wims19a} experimentally
investigated efficient bit erasure in a nanoscale flux qubit device. We found a
natural partition of trajectories into classes that served well both to
strengthen the Second Law and to dissect the entire process work distribution.
The latter identified simple components characterizing the full work
distribution's features---features functionally critical to efficient bit
erasure.

Helpfully, the TCFT can be derived under less severe restrictions on the
system than Sec. \ref{sec:model} assumed. For example, the nature of the
external influence that we called the control device can be allowed to
instantiate nonconservative forces on the system. Moreover, the thermal
environment does not need to be fixed at inverse temperature $\beta$ for the
duration of the protocol. That is, we need not require that the steady state of
the system is in equilibrium nor that all forces acting on the system are
conservative. Relaxing other assumptions is possible too. The essential
equality needed for a version of the TCFT to also hold is one like the DFT:
\begin{align*}
  \frac{\forP(\traj)}{\revP(\traj)}=g(\traj)
  ~,
\end{align*}
where $g$ is some function of trajectory. Then a version of the TCFT holds
where the exponential entropy difference $e^{-\Entlike}$ is replaced with the
function $g$. This follows by the same logic as presented in Sec.
\ref{sec:tcft}. However, how such a generalization of our presentation can be
used remains to be explored.

\section{Conclusions}
\label{sec:conclusion}

We presented the TCFT's core theory. With it, we detailed a path to solving for
free-energy differences more efficiently than before. We also showed how to
strengthen the Second Law in the presence of impoverished knowledge of any
nonequilibrium and dissipative process. This led to a suite of new results that
further advanced our understanding of how fluctuations underpin nonequilibrium
thermodynamics.

We also showed how the TCFT fits more broadly within the ranks of fluctuation
theorems. It unifies many previously known, but distinct results, spans the
detailed and integral fluctuation theorems, and is rooted in the same
conceptual foundation of time symmetries on small dynamical systems.

Follow-on efforts will expand on the way that the TCFT breaks free energy
estimation from the tyranny of rare events. This will also clarify the role of
metastable free energies in describing experimentally inaccessible free
energies of interest and related thermodynamic costs. Via explicit examples,
this will also showcase how the TCFT solves for these metastable free energies.

\section{Acknowledgments}

We thank Jacob Hastings, Kyle Ray, Paul Riechers, and Mikhael Semaan for
helpful discussions. The authors thank the Telluride Science Research Center
for hospitality during visits and the participants of the Information Engines
Workshops there. JPC acknowledges the kind hospitality of the Santa Fe
Institute, Institute for Advanced Study at the University of Amsterdam, and
California Institute of Technology. This material is based upon work supported
by, or in part by, FQXi Grant number FQXi-RFP-IPW-1902 and U.S. Army Research
Laboratory and the U.S.  Army Research Office under grants W911NF-21-1-0048 and
W911NF-18-1-0028.

{\bf Data Availability Statement:} Neither data nor programming code
is necessary to support this work's results.

\appendix

\section{Alternative TCFT Derivations}
\label{app:AltTCFTDerivations}

Equation (\ref{eq:exptcft})'s Exponential Class Fluctuation Theorem
(\exptcftabbr\,) can be derived from at least two prior results---from path
ensemble averaging and from a master fluctuation theorem.

\subsection{Path Ensemble Average}

We first give a derivation from a generalization of Crooks' Path Ensemble
Average to arbitrary privileged distributions. The Path Ensemble Average
result is expressed in Eq. (15) of Ref. \cite{Croo00a}:
\begin{align*}
\langle \mathcal{F} e^{-\Entlike(\traj)} \rangle_F
= \langle \widehat{\mathcal{F}} \rangle_{R'}
  ~.
\end{align*}
Here, $\mathcal{F}$ is an arbitrary trajectory functional,
$\widehat{\mathcal{F}}$ is its time reverse, defined by
$\widehat{\mathcal{F}}(\traj) = \mathcal{F}(\revtraj)$, and $\langle \cdot
\rangle_{y}$ denotes a trajectory ensemble average over the forward ($y=F$) or
\physrevname\, of the reverse ($y=R'$) processes.

We first convert to the formal reverse representation for convenience:
\begin{align*}
  \avgat{\widehat{\mathcal{F}}}{R'}
  &= \int d\traj \physrevP(\traj) \mathcal F(\rev\traj) \\
  &= \int d\traj \revP(\rev\traj) \mathcal F(\rev\traj) \\
  &= \int d\traj \revP(\traj) \mathcal F(\traj) \\
  &= \avgat{\mathcal F}{R}
  ~,
\end{align*}
where $R$ denotes that the average is taken over the \formrevname\, of the
reverse process.  This gives:
\begin{align}
\langle \mathcal{F} e^{-\Entlike(\traj)} \rangle_F
  = \langle {\mathcal{F}} \rangle_R
  ~.
\label{eq:crooks_ensemble_avg_trans}
\end{align}

Consider an arbitrary trajectory class $\class \in \classes$. Then let
$\mathcal{F}(\traj) = [ \traj \in \class ]$---$\class$'s characteristic function---for all $\traj \in \trajs$. Then the LHS of Eq.
(\ref{eq:crooks_ensemble_avg_trans}) becomes:
\begin{align*}
  &\int d\traj \forP(\traj) [\traj \in \class ] e^{-\Entlike(\traj)} \\
  &\quad\quad= \int_\class d\traj \forP(\traj | \class)
  \forP(\class) e^{-\Entlike(\traj)} \\
  &\quad\quad= \forP(\class) \int_\class d\traj \forP(\traj | \class)
  e^{-\Entlike(\traj)}\\
  &\quad\quad= \forP(\class) \avgatclass{e^{-\Entlike}}
  ~.
\end{align*}
Equation (\ref{eq:crooks_ensemble_avg_trans})'s RHS is simply $\revP(\class)$.
Combining yields Eq. (\ref{eq:exptcft}), showing that the \exptcftfull\, is
the path or trajectory ensemble average of a characteristic function
$[\traj \in \class]$.

\subsection{Master Fluctuation Theorem}

For Langevin dynamics, invoke Ref. \cite{Seif12a}'s Master Fluctuation
Theorem:
\begin{align*}
\avgat{g(\{S_\alpha\}) e^{-\Entlike}}{F}
&=\avgat{g(\{\epsilon_\alpha S_\alpha^\dagger)\}}
  {R'}
  ~.
\end{align*}
This is Eq. (78) there.
$\{S_\alpha\}$ is a set of functions of the system microstate trajectories for
the forward process. $\{S_\alpha^\dagger\}$ is a corresponding set of functions
of the trajectories for the reverse process with the following relationship:
$S_\alpha^\dagger(\rev\traj) = \epsilon_\alpha S_\alpha(\traj)$,
where $\epsilon_\alpha = \pm 1$.
And, $g$ is any function of the set $\{S_\alpha\}$.

To derive the \exptcftabbr, we consider the singleton
$\{S_\alpha(\traj)\} = \{[\traj\in\class]\}$.
We then define $S_\alpha^\dagger(\traj) = [\rev\traj \in \class]$,
giving
$S_\alpha^\dagger(\rev\traj) = [\traj \in \class] = S_\alpha(\traj)$
and $\epsilon_\alpha = 1$.
Then, set $g$ to be the identity, obtaining:
\begin{align*}
  \avgat{[\traj \in \class] e^{-\Entlike}}{F}
  &= \avgat{[\rev\traj \in \class]}{R'}
  ~.
\end{align*}
This is now in the form of Crooks' Path Ensemble Average for
$\mathcal{F}(\traj) = [\traj \in \class]$.

\section{Irreversibility as Entropy-Difference Variability}
\label{sec:app_classirr}

The following shows that the average class irreversibility $\avgirratclass$
tracks the variability of entropy difference when small. Moreover,
$\avgirratclass$ and variability both necessarily go to zero together. And so,
the TCFT shows that finding a class with a narrow entropy-difference
distribution is tantamount to minimizing the class irreversibility and
class-average entropy difference.

First, translate $\Entlike$ into $x=\Entlike - \langle \Entlike
\rangle_\class$, its difference from its average:
\begin{align*}
\langle e^{-\Entlike} \rangle_\class
= \langle e^{-x} \rangle_\class \, e^{-\langle \Entlike \rangle_\class}
 ~.
\end{align*}

Then, Taylor expand:
\begin{align*}
\langle e^{-x} \rangle_\class
&= \sum_{n=0}^{\infty} \frac{(-1)^n}{n!} \langle x^n \rangle_\class \\
&= 1 + a
 ~,
\end{align*}
with:
\begin{align*}
a & \equiv \sum_{n=2}^\infty \frac{(-1)^n}{n!} \langle x^n \rangle_\class \\
  & \geq 0
~.
\end{align*}
When $\Entlike$'s variability is small, the $x$ are typically small and the
second order term $\avgatclass{x^2}$ dominates in $a$. $a$ is, then, the
variance of $\Entlike$ over $\class$.

Then, using Eqs. (\ref{eq:exptcft}) and (\ref{eq:entlike_classdecomp}), we
have:
\begin{align*}
e^{-\classsurpatclass} = (1+a) e^{-\classsurpatclass -\avgirratclass}
~.
\end{align*}
This gives:
\begin{align*}
  \avgirratclass = \ln(1+a)
  ~.
\end{align*}
Since $a$ goes as the variance, $\avgirratclass$ is also a measure of
$\Entlike$'s variability in $\class$ in the small variability limit.

\section{Free Energy Estimate Distribution}
\label{app:Free Energy Estimate Distribution}

If we wish to estimate the change in free energy from the work distributions
of a collection of forward and reverse experiments that start in equilibrium, the TCFT provides a relation for the free-energy differences of each trajectory
class $C$:
\begin{align*}
e^{-\Delta f_C }=  \langle e^{-\beta^{-1}W} \rangle_C  \frac{P(C)}{R(C)},
\end{align*}
which each equal the change in free energy $\Delta F^\text{eq}=\Delta f_C$.
Let us consider a particular experiment with two energy levels that start at:
\begin{align*}
E_0(A) & =- \beta^{-1}\ln (1- \epsilon) 
\\ E_0(B) & =-\beta^{-1} \ln \epsilon,
\end{align*}
with the corresponding equilibrium distribution:
\begin{align*}
\pi_0(A)& =1-\epsilon
\\ \pi_0(B) & = \epsilon.
\end{align*}
Note that, because $E_t(s) \equiv F^\text{eq}_t - \beta^{-1} \ln \pi_t(s)$,
the free energy in this case is zero initially: $F^\text{eq}_0=0$.
We then change the energy level instantaneously to the final energy landscape:
\begin{align*}
E_\tau(A)=E_\tau(B) & = - \beta^{-1} \ln \pi_\tau(A) \\
  & =- \beta^{-1}\ln \pi_\tau(B) \\
  & = \beta^{-1}\ln 2,
\end{align*}
which also has zero free energy $F^\text{eq}_\tau =0$. If the initial state is
$s$, then it remains $s$ and the work investment is:
\begin{align*}
W(A) & =E_\tau(A)-E_0(A) \\
	& =\beta^{-1}\ln (2(1-\epsilon)) \\
W(B) & =E_\tau(B)-E_0(B) \\
	& =\beta^{-1}\ln (2\epsilon)
  ~.
\end{align*}

To evaluate the free-energy difference estimate, we note
$\Delta \tilde{f}_C$ is itself a function of the estimated probability of
realizations of the experiment that starts in $A$:
\begin{widetext}
\begin{align*}
\Delta \tilde{f}_C(\tilde{P}(A),\tilde{R}(A))  &
=  - \beta^{-1} \ln \left(  \frac{ \delta_{A \in C} \tilde{P}(A) e^{-W(A)}
 + \delta_{B \in C} (1-\tilde{P}(A)) e^{-W(B)} }{ \delta_{A \in C} \tilde{P}(A)
 + \delta_{B \in C} (1-\tilde{P}(A))} \frac{ \delta_{A \in C} \tilde{P}(A)
 + \delta_{B \in C} (1-\tilde{P}(A))  }{ \delta_{A \in C} \tilde{R}(A)
 + \delta_{B \in C} (1-\tilde{R}(A))} \right)
\\ & = - \beta^{-1} \ln \left(  \frac{ \delta_{A \in C} \tilde{P}(A) e^{-W(A)}
 + \delta_{B \in C} (1-\tilde{P}(A)) e^{-W(B)} }{ \delta_{A \in C} \tilde{R}(A)
 + \delta_{B \in C} (1-\tilde{R}(A))} \right)
 ~.
\end{align*}
We use frequentist statistics to estimate the probabilities of our initial
states and resultant works. Given $N$ forward experiments and $N$ reverse
experiments, the probability of realizing free energy $\Delta F$ is determined
by evaluating the number $n_A$ of times the forward experiment starts in $A$
and the number $n_A^R$ of times the reverse experiment starts in $A$:
\begin{align*}
\Pr(\Delta \tilde{f}_C=\Delta F) & = \sum_{n_A,n^R_A}\Pr(\Delta \tilde{f}_C
= \Delta F,\tilde{P}(A)=n_A/N,\tilde{R}(A)=n_A^R/N) 
\\ & = \sum_{n_A,n^R_A} \delta_{\Delta F ,\Delta \tilde{f}_C(n_A/N,n_A^R/N)}
 \Pr(\tilde{P}(A)=n_A/N,\tilde{R}(A)=n_A^R/N)
 ~.
\end{align*}
\end{widetext}

For $N$ experiments, we can combinatorially evaluate the probability of
realizing $n_A$ and $n_A^R$ as a function of $N$:
\begin{align*}
\Pr(\tilde{P}(A)=n_A/N)= (1-\epsilon)^{n_A}\epsilon^{N-n_A} {N \choose n_A}
\end{align*}
and:
\begin{align*}
\Pr(\tilde{R}(A)=n^R_A/N)= 2^{-N}{N \choose n^R_A}
	~.
\end{align*}
Assuming that the forward and reverse experiments are performed independently,
the joint probability of realizing $n_A$ and $n_A^R$ is:
\begin{align*}
\Pr(\tilde{P}(A) & =n_A/N,\tilde{R}(A)=n_A^R/N) \\
 & = \Pr(\tilde{R}(A)=n^R_A/N)\Pr(\tilde{P}(A)=n_A/N) \\
 & =(1-\epsilon)^{n_A}\epsilon^{n_A-N} {N \choose n_A}2^{-N}{N \choose n^R_A}. \nonumber
\end{align*}
We then compute the probability of our free energy estimate
$\Pr(\Delta \tilde{f}_C=\Delta F)$ for $N$ experiments with
$\pi_0(A)=\epsilon$:
\begin{align*}
\Pr(\Delta \tilde{f}_C & =\Delta F)  \\
& = \sum_{n_A,n^R_A} \delta_{\Delta F ,\Delta \tilde{f}_C(n_A/N,n_A^R/N)} \\
 & \qquad \times (1-\epsilon)^{n_A}\epsilon^{n_A-N} {N \choose n_A}2^{-N}{N \choose n^R_A}
   ~. 
\end{align*}

\end{document}